\documentclass[aps,prc,preprint,showpacs,groupedaddress,floatfix,amsmath,amssymb]{revtex4-1}
\usepackage{color}
\usepackage{graphicx}
\usepackage{CJK}
\usepackage{longtable,booktabs}
\usepackage{ltablex}
\usepackage{subfigure}
\usepackage{multirow}
\usepackage{makecell}
\usepackage{graphicx}
\newcommand{\red}[1]{\textcolor{red}{{#1}}}

\begin{document}
\title{Influence of moments of inertia on transverse wobbling mode in odd-mass nuclei}
\author{H. Zhang}
\affiliation{Shandong Provincial Key Laboratory of Optical Astronomy
and Solar-Terrestrial Environment, Institute of Space Sciences, Shandong University, Weihai 264209, People's Republic of China}
\author{B. Qi}\email[]{bqi@sdu.edu.cn}\thanks{}
\affiliation{Shandong Provincial Key Laboratory of Optical Astronomy
and Solar-Terrestrial Environment, Institute of Space Sciences, Shandong University, Weihai 264209, People's Republic of China}
\author{X. D. Wang}
\affiliation{Shandong Provincial Key Laboratory of Optical Astronomy
and Solar-Terrestrial Environment, Institute of Space Sciences, Shandong University, Weihai 264209, People's Republic of China}
\author{H. Jia}
\affiliation{Shandong Provincial Key Laboratory of Optical Astronomy
and Solar-Terrestrial Environment, Institute of Space Sciences, Shandong University, Weihai 264209, People's Republic of China}
\author{S. Y. Wang}
\affiliation{Shandong Provincial Key Laboratory of Optical Astronomy
and Solar-Terrestrial Environment, Institute of Space Sciences, Shandong University, Weihai 264209, People's Republic of China}

\begin{abstract}
The reported transverse wobbling band in odd-mass $^{105}$Pd has been reinvestigated
by the triaxial particle rotor model. Employing different parameter sets of moment of inertia (MOI),  several calculated results could be in good agreement with the experimental data, which show
distinct modes of rotational excitation, respectively.
These modes are sensitive to the ratio
between the MOI at intermediate and short axis. With the increase of this ratio, a wobble about the short axis of the total angular momentum  is gradually changed  to a wobble about the intermediate axis.
In addition, it is exhibited that
precession and tunneling are two aspects of the quantum wobbling motion. The tunneling aspect dominates
in the yrare states of $^{105}$Pd.
The present results in $^{105}$Pd show the complexity of the transverse wobbling mode.
\end{abstract}

\maketitle

\section{Introduction}

The concept of wobbling motion in atomic nuclei was originally
introduced by Bohr and Mottelson in the 1970s~\cite{Bohr75}. It is described as the small amplitude harmonic oscillations of the total angular
momentum with respect to the principal axis having the
largest moment of inertia (MOI)~\cite{Bohr75}. With increasing angular momentum components along the two axes with smaller MOI, it
excites a family of $\Delta I=2$ rotational bands corresponding to increasing phonon quanta $n$.
These bands are connected with enhanced collective $\Delta I=1$ $E2$ transitions. The pioneering work of  the
microscopic description on wobbling motion  was introduced by Marshalek~\cite{Marshalek79}.
It was first proposed to appear in triaxial even-even nuclei with
zero quasiparticle configuration~\cite{Bohr75}, while the corresponding experimental evidence is still scarce.
For instance, the $\gamma$ bands in $^{112}$Ru were interpreted as a possible evidence~\cite{Ru112}, but no interband electromagnetic transition rates were reported.

Instead, wobbling bands have been widely reported in odd-mass nuclei.
It was first suggested in the triaxial strongly deformed bands of $^{163}$Lu~\cite{Lu1631,Lu1632}
and later in other neighboring nuclei $^{161}$Lu~\cite{Lu161}, $^{165}$Lu~\cite{Lu165}, $^{167}$Lu~\cite{Lu167}, $^{167}$Ta~\cite{Ta167}.
Then, experimental evidences of wobbling motion was also reported in other normal deformed nuclei: $^{105}$Pd~\cite{Pd105}, $^{130}$Ba~\cite{Ba130,Ba1302}, $^{133}$Ba~\cite{Ba133}, $^{135}$Pr~\cite{Pr135,Pr1352}, and $^{183}$Au~\cite{Au183}, where the wobbling energies decrease with increasing spin contrary to the theoretical expectations in Ref.~\cite{Bohr75}. Frauendorf and D\"{o}nau~\cite{Frauendorf14} interpreted this behavior as the odd particle aligned perpendicular to the principal axis with the
largest MOI, which  was defined as transverse wobbling (TW). In addition,
$^{127}$Xe~\cite{Xe127}, $^{133}$La~\cite{La133} and $^{187}$Au~\cite{Au187} exhibit longitudinal wobbling (LW), where the odd particle aligned parallel to the principal axis with the
largest MOI. For the ideal TW (LW), the total angular momentum is aligned along the short (intermediate) axis in the yrast band. With one phonon excitation, the precession of total angular momentum occurs with respect to the short (intermediate) axis~\cite{Frauendorf14,Au187}, which is presented
pictorially in Figs.~\ref{fig:picture}(a) and \ref{fig:picture}(c). Very recently, an experiment for $^{135}$Pr provided the crucial evidence against the proposed wobbling nature~\cite{Pr135tilted}, which questioned the interpretation of wobbling.

On the theoretical side, the wobbling excitation got extensive descriptions with the particle rotor model (PRM)~\cite{Frauendorf14,Hamamoto02,
Hamamoto2003PRC, Hagemann04, Streck18,Broocks21,CQB19} and
its approximation solutions~\cite{Tanabe08,Tanabe10,Tanabe17, Budaca2018PRC,Budaca2021, Raduta2018, Raduta2020JPG, Raduta2020PRC,Raduta2021,Lawrie20}.
In addition,
the random phase approximation~\cite{Matsuzaki02, Matsuzaki2003PRC,Matsuzaki2004PRC, Matsuzaki2004PRC_v1,Shimizu2005PRC,Shimizu2008PRC,Shoji09, Frauendorf2015PRC, Nakatsukasa16},
the angular momentum projection  method~\cite{Pr1352,Ba1302,Shimada18},
and the collective Hamiltonian method~\cite{CHENQB14, CHENQB16,WXH18} are used to discuss this issue.
However, it should be noted that there are still increasingly loud debates on the TW in odd-mass nuclei.
Reference~\cite{Tanabe17} used an approximation to the PRM to study the stability of TW and concluded there is an absence of the wobbling mode around the axis with medium MOI.
This issue was
further discussed~\cite{Budaca2018PRC,Frauendorf182,Tanabe18}.
Furthermore, an alternative mode for the candidate wobbling band was proposed in Ref.~\cite{Lawrie20}, $i.e.$, a precession of the total angular momentum around a certain tilt axis (called tilted precession bands), which is presented in Fig.~\ref{fig:picture}(b).

Whether the stable TW exist in odd-mass nuclei is still an open question. In this paper, we took $^{105}$Pd as an example and made an attempt to clarify some disputes about TW motion. Recently,
two bands built on $\nu h_{11/2}$ configuration in $^{105}$Pd had been suggested as TW due to
the $\Delta I=1$ transitions between them with a predominant $E2$ component and
the decreasing wobbling energy with increasing spin~\cite{Pd105}.
The behavior of the wobbling bands in $^{105}$Pd has been reinvestigated by the quantum PRM. The different parameter sets of MOI will be searched to reproduce the experimental data. The corresponding rotational modes will be analyzed in detail in terms of the azimuthal plot of angular momentum. Consequently, the picture of TW will be examined.

\section{Theoretical Framework}

The Hamiltonian of the particle rotor model can be expressed as~\cite{Bohr75,Ring}
\begin{eqnarray}
\hat{H}=\hat{H}_\textrm{{coll}}+\hat{H}_\textrm{{intr}}
\end{eqnarray}
with the collective Hamiltonian
\begin{eqnarray}
\begin{aligned}
\hat{H}_\textrm{{coll}}&=\sum_{k=1}^{3}\frac{\hat{R}_{k}^{2}}{2{\cal J}_{k}}=\sum_{k=1}^{3}\frac{(\hat{I}_{k}-\hat{j}_{k})^{2}}{2{\cal J}_{k}}\\
&=\sum_{k=1}^{3}\frac{\hat{I}_{k}^{2}}{2{\cal J}_{k}}+\frac{\hat{j}_{k}^{2}}{2{\cal J}_{k}}-\frac{\hat{I}_{k}\hat{j}_{k}}{{\cal J}_{k}},
\end{aligned}
\end{eqnarray}
where $k=1,2,3$  denotes the three principal axes of the body-fixed frame corresponding to the intermediate ($m$), short ($s$), and long ($l$) axis for $0^{\circ}<\gamma<60^{\circ}$, respectively.  Here $\hat{R}_{k}$, $\hat{I}_{k}$, and $\hat{j}_{k}$, respectively, denote the angular momentum operators for the core and nucleus as well as the valence nucleon.
The parameters ${\cal J}_{k}$ are the MOI for three principal axes.

Two models of MOI are usually taken in the study of wobbling~\cite{Tanabe17,Tanabe18}, $i.e.$, the hydrodynamical MOI:
\begin{eqnarray}
{\cal J}_{k}={\cal J}_0 \sin ^{2}(\gamma-\frac{2 k \pi}{3})
\end{eqnarray}
and the rigid-body MOI:
\begin{eqnarray}
{\cal J}_{k}= \frac{{\cal J}_0}{1+(\frac{5}{16\pi})^{1/2} \beta }[1-(\frac{5}{4\pi})^{1/2}\beta \cos (\gamma+\frac{2 k \pi}{3})].
\end{eqnarray}
The above expressions are different from the standard convention of $k = x, y, z$, in which the sign of
$\gamma$ is meaningful~\cite{Hamamoto02,Ring}.

The intrinsic Hamiltonian $\hat{H}_\textrm{{intr}}$ describes a single valence nucleon in a high-$j$ shell
\begin{eqnarray}
\hat{H}_\textrm{{intr}}=\pm\frac{1}{2} C\left\{\cos \gamma\left(j_{3}^{2}-\frac{j(j+1)}{3}\right)+\frac{\sin \gamma}{2 \sqrt{3}}\left(j_{+}^{2}+j_{-}^{2}\right)\right\},
\end{eqnarray}
where $\pm$ refers to a particle or a hole state and the coefficient $C$ is proportional to the quadrupole deformation parameter $\beta$~\cite{Qi09}. Differing from the frozen approximation proposed in Ref.~\cite{Frauendorf14}, we present our calculations with free odd-particle angular momentum.

To obtain the PRM solutions, the total Hamiltonian must be diagonalized in a complete basis space, which couples the rotation of the core with the intrinsic wave function of a valence nucleon. When pairing correlations are neglected, one can construct the so-called strong coupling basis~\cite{Ring,Qi09},
\begin{eqnarray}
|I M K j \Omega +\rangle= \sqrt{\frac{2 I+1}{16 \pi^{2}}}\left[D_{M K}^{I}|j \Omega\rangle +(-1)^{I-j} D^{I}_{M-K}|j-\Omega\rangle\right],
\end{eqnarray}
where $I$ denotes the total angular momentum of the odd-mass nuclei and $K$ refers to the projection onto the 3-axis of the intrinsic frame. Furthermore, $\Omega$ is the 3-axis component of the valence nucleon angular momentum $j$ in the intrinsic frame. Under the requirement of the $D_{2}$
symmetry of a triaxial nucleus,
$K-\Omega$ need to be an even integer. The matrix elements of total Hamiltonian can be evaluated in this basis, and the diagonalization gives eigenenergies and eigenstates for the PRM Hamiltonian. The wave function of PRM can be expressed as
\begin{eqnarray}\label{PRMfun}
|I M \rangle = \sum\limits_{K,\Omega} C_{IK j\Omega} |I M K j \Omega +\rangle.
\end{eqnarray}

The reduced  electric quadrupole transition probabilities are calculated by the operator
\begin{eqnarray}
{\hat{\cal{M}}}(E2, \mu) =\sqrt{\frac{5}{16\pi}}\hat{Q}_{2\mu}
\end{eqnarray}
with the obtained wave functions. The quadrupole moments
in the laboratory frame ($\hat{Q}_{2\mu}$) and the intrinsic
system ($\hat{Q}'_{2\mu}$) are connected by the relation
\begin{eqnarray}\label{Q0}
\begin{aligned}
\hat{Q}_{2\mu}&=\mathcal{D}_{\mu 0}^{2 *} \hat{Q}_{20}^{\prime}
+\left(\mathcal{D}_{\mu 2}^{2 *}+\mathcal{D}_{\mu -2}^{2 *}\right) \hat{Q}_{22}^{\prime}\\
&=\mathcal{D}_{\mu 0}^{2 *} Q\cos \gamma
+\left(\mathcal{D}_{\mu 2}^{2 *}+\mathcal{D}_{\mu -2}^{2 *}\right) \frac{1}{\sqrt{2}}Q\sin \gamma.
\end{aligned}
\end{eqnarray}

For $M1$ transitions, the magnetic dipole transition operator can be written as
\begin{eqnarray}
 \hat{\cal{M}} (M1,\mu) = \sqrt{\frac{3}{4\pi}} \frac{e\hbar}{2Mc}
 [(g_j-g_R)\hat{j}_{\mu}],
\end{eqnarray}
where $g_j$ and $g_{R}$ are, respectively, the effective gyromagnetic ratios for valence nucleon and the collective core, and $\hat{j}_{\mu}$ denotes the spherical tensor in the laboratory frame.

Using wave functions obtained from the PRM, one can calculate the
expectation values of the core angular momentum as
\begin{eqnarray}
\langle R_{k}^2 \rangle^{1/2} & = \langle I M |(\hat{I_{k}}-\hat{j_{k}})^{2}|I M \rangle^{1/2}.
\end{eqnarray}

To illustrate clearly the angular-momentum geometry,  the probability distribution of the total angular momentum on the $(\theta, \varphi)$
plane, $i.e.$, azimuthal plot~\cite{Streck18,CFQ17,CQB18,CQB19}, is calculated.
Here $(\theta, \varphi)$ are the orientation angles of the angular-momentum
vector $\textit{\textbf{I}}$ (expectation value with $M = I$ ) with
respect to the intrinsic frame. The polar angle $\theta$ is the
angle between $\textit{\textbf{I}}$ and the $l$ axis, and the
azimuthal angle $\varphi$ is the angle between the projection of $\textit{\textbf{I}}$ on
the $sm$ plane and the $m$ axis. The
profile is calculated as~\cite{Streck18,Qi21}
\begin{eqnarray}\label{APlot}
\begin{aligned}
\mathcal{P}^{(\nu)}(\theta, \varphi)&=\langle I, \theta \varphi \mid I I \nu\rangle^{2}\\
&=\frac{2I+1}{8\pi}\sum\limits_{K K^{\prime}} D_{K I}^{I *}(\theta, \varphi, 0)
\rho_{KK'}^{(\nu)} D_{K^{\prime} I}^{I}(\theta, \varphi, 0),
\end{aligned}
\end{eqnarray}
where $\rho_{KK'}^{(\nu)}=\sum\limits_{\Omega}C_{IKj\Omega}^{(\nu)}C_{IK'j\Omega}^{(\nu)}$ with the expansion coefficients $C_{IKj\Omega}^{(\nu)}$ in Eq.~(\ref{PRMfun}).

Here we further calculate the profile for the valence nucleon angular momentum~\cite{Frauendorf20},
\begin{eqnarray}
\begin{aligned}
\mathcal{P}^{(\nu)}(\theta, \varphi)&=\langle j, \theta \varphi \mid j j\nu\rangle^{2}\\
&=\frac{2j+1}{8\pi}\sum\limits_{\Omega \Omega^{\prime}} D_{\Omega j}^{j *}(\theta, \varphi, 0)
\rho_{\Omega \Omega^{\prime}}^{(\nu)} D_{\Omega^{\prime}j}^{j}(\theta, \varphi, 0)
\end{aligned}
\end{eqnarray}
with the density matrix
$\rho_{\Omega\Omega^{\prime}}^{(\nu)}=\sum\limits_{K} C_{I K j\Omega}^{(\nu)} C_{I K j\Omega^{\prime}}^{(\nu)}$.

Both the profiles $\mathcal{P}^{(\nu)}(\theta, \varphi)$ fulfill the normalization condition
$$\int_{0}^{\pi} d \theta \sin \theta \int_{0}^{2 \pi} d \varphi \mathcal{P}^{(\nu)}(\theta, \varphi)=1.$$

\section{Results and Discussion}


\subsection{Description of the data}

Following Ref.~\cite{Pd105}, the configuration $\nu (1h_{11/2})^{1}$
with a triaxial shape of $\beta=0.27$ and $\gamma=25^{\circ}$ for $^{105}$Pd is taken in our calculation.
The triaxial rotor is parametrized by three angular-momentum-dependent MOI
 ${\cal J}_{k}=a_{k}\sqrt{1+bI(I+1)}$~\cite{Pd105}.
Both of hydrodynamical and rigid MOI with such triaxial shape are ${\cal J}_{1}>{\cal J}_{2}>{\cal J}_{3}$, $i.e.$, ${\cal J}_{m}>{\cal J}_{s}>{\cal J}_{l}$. In the following discussion, the three principal axes  are marked by the suffixes $m$,  $s$, and $l$  for  convenience.
For $\gamma=25^{\circ}$, the ratio between the MOI at intermediate and short axis ${\cal J}_{m}/{\cal J}_{s}$ is approximately equal to $1:0.9$ for a rigid-body MOI and $1:0.3$ for a hydrodynamical MOI.
Four typical sets of parameters are adopted and listed in Table~\ref{tab:Pd105MOIset}, in which ${\cal J}_{m}/{\cal J}_{s}$ takes values of  $1:0.9$~(A), $1:0.7$~(B), $1:0.5$~(C), and $1:0.3$~(D), respectively. The corresponding parameters $a_m$ and $b$
are determined by fitting the energy spectra of yrast band A and yrare band B in $^{105}$Pd.
In Ref.~\cite{Pd105}, ${\cal J}_{m}/{\cal J}_{s}=1:0.63$ with $b=0.023$ was adopted, which is between the ratio of the parameter sets (B) and (C).
${\cal J}_{l}$ is determined by fitting the energy difference between yrast and yrare band, and such an energy difference is found to be smaller for the bigger value of ${\cal J}_{l}$.

For the electromagnetic transitions, we used the intrinsic
quadrupole moments $Q=(3/\sqrt{5\pi})R_{0}^{2}Z\beta=3.0$ eb,
the gyromagnetic ratios $g_{R}=Z/A=0.438$ for the rotor and $g_{\nu }(h_{11/2})=-0.209$ for the neutron.
Note that a quenching factor of $0.36$ for $g$ is introduced in our calculation same as Ref.~\cite{Pd105}. This is due to the scissor mode which is mixed with the wobbling motion and cannot be considered in the PRM calculations~\cite{Pd105}.

The calculated energy spectra, wobbling energies and reduced transition probability ratios [$i.e.$, $B(E2)_{\textrm{out}}/B(E2)_{\textrm{in}}$ and $B(M1)/B(E2)_{\textrm{in}}$]  in comparison with the experimental data are shown in Fig.~\ref{fig:reproduce}.
The four adopted parameter sets of the MOI in Table~\ref{tab:Pd105MOIset} and the corresponding results in Fig.~\ref{fig:reproduce} are denoted as (A), (B), (C), and (D), respectively.
For the energy spectra, the results of four sets of parameters are all in good agreement with the experimental data. The wobbling energies obtained by four parameter sets can reproduce the decreasing trend with spin from $I=13/2\hbar$ to $29/2\hbar$. The increasing trend of wobbling energies in the region $I \geq 33/2\hbar$ was attributed to three quasiparticle configuration in Ref.~\cite{Pd105}. Here the trends of wobbling energy at the whole spin region can also be described with one quasiparticle configuration adopted parameters (C) and (D).

The $\Delta I=1$ transitions connecting the wobbling bands should be dominated by an $E2$ component, due to the collective motion of the entire nuclear charge.
The strong $E$2 component of such transitions in $^{105}$Pd and the reduced transition probability ratios $B(E2)_{\textrm{out}}/B(E2)_{\textrm{in}}$ and $B(M1)/B(E2)_{\textrm{in}}$ are reproduced by the present calculation with the four sets of parameters. $B(E2)_{\textrm{out}}/B(E2)_{\textrm{in}}$ depends on $Q^{'}_{20}$ and $Q^{'}_{22}$ in Eq.~(\ref{Q0}), which is associated with $\gamma$ values.  It is found that $B(E2)_{\textrm{out}}/B(E2)_{\textrm{in}}$ will be significantly underestimated if a small $\gamma$ parameter is adopted in the calculation.

\subsection{Realization of transverse wobbling}

In the lowest row of  Fig.~\ref{fig:reproduce}, the expectation values of the core angular momentum components along the $m$, $s$, and $l$ axes are plotted. For parameter set (A), the value of ${\cal J}_m$ is close to ${\cal J}_s$, which agrees with the rigid-body model. For the yrast band, the core angular momentum increases along the $s$ axis more than that along the $m$ axis. For the yrare band, the $s$ and $m$ components of the core angular momentum are about the same. This is consistent with a tilt of core angular momenta about the $s$ axis.
For the parameter set (D), where the ratio ${\cal J}_m/{\cal J}_s=1:0.3$ is closed to that of hydrodynamical model, the component of the core angular momentum  is mainly on the $m$ axis for yrast and yrare bands. For the parameter sets (B) and (C), the angular momentum orientation with the minimal energy gradually transfers from the $s$ axis to the $m$ axis.

These different orientations of angular momentum are driven by the competition between the Coriolis term and rotational term to minimize energy.
As the valence nucleon angular momentum is mainly along the $s$ axis, the Coriolis term $-\hat{I}_{k}\hat{j}_{k}/{\cal J}_{k}$ contributes the most (least) to the energy  when the total angular momentum is along the $s$ ($m$) axis.
For the rotational term $\hat{I}_{k}^{2}/2{\cal J}_{k}$, it is largest (smallest) when the total angular momentum is along the $s$ ($m$) axis.
The MOI has a great influence on the orientation of core angular momentum.

To further illustrate the angular momentum geometry of
the wobbling motion, a probability profile on the $(\theta, \varphi)$
plane, $i.e.$, azimuthal plot~\cite{Streck18,CFQ17,CQB18,CQB19}, is provided in the following for both the angular momentum of nucleus and the valence nucleon.

In Figs.~\ref{fig:Aplota}-\ref{fig:Aplotd}, the obtained profiles $\mathcal{P}(\theta, \varphi)$ are shown at $I=11.5, 15.5$, and $19.5 \hbar$ for the yrast band and at $I=12.5,16.5$, and $20.5 \hbar$ for the yrare band with the four parameter sets of MOI in Table~\ref{tab:Pd105MOIset}. The distributions $\mathcal{P}(\theta, \varphi)$ of angular momentum are always concentrated around $\theta=90^{\circ}$.
This is because the angular momentum is prone to locate in the $sm$ plane to obtain the lowest energy.
To make it more visualized, the probability distribution of angular momenta in the $sm$ plane ($\theta=90^{\circ}$) are shown in the lower panels of Figs.~\ref{fig:Aplota}-\ref{fig:Aplotd}.
The radial coordinate represents the magnitude of angular momentum ranging from 0 to 21$\hbar$, and the angle coordinate $\varphi$ from $0^{\circ}$ to $360^{\circ}$.
The corresponding probabilities in the upper and lower panels have the same color schemes.

Adopted the different parameter sets of the MOI in Table~\ref{tab:Pd105MOIset}, distinct patterns of angular momentum are obtained in the calculations.
In Fig.~\ref{fig:Aplota}, the $\mathcal{P}(\theta, \varphi)$ results with parameter set (A) are shown, where the MOI is close to the rigid-body model. The maximum of $\mathcal{P}(\theta, \varphi)$ of the total and valence nucleon angular momentum are always located at $\varphi=90^{\circ}$, namely along the $s$ axis for $I=11.5\hbar, 15.5\hbar, 19.5\hbar$ in the yrast band.
For the yrare band, the total angular momentum has the maximal probabilities lying symmetrically with respect to the $s$ axis. While the angular momentum of valence nucleon still locates on the $s$ axis.
Combined with the orientation of the two,
it is straightforward to indicate the geometry of core angular momentum.
The core angular momentum aligns along the $s$ axis in the yrast band,
the motion in the yrare band corresponds to the oscillation (wobble) between two symmetrically orientations of angular
momentum about the $s$ axis.
Similar azimuthal plots  had been obtained in previous works, $e.g.$, Refs.~\cite{CQB19,Streck18,Ba130}, which were regarded as TW motion. They mentioned that the maximal probabilities of yrare states lying on a rim around the minimum reflects the wobbling motion (or precession) in this way~\cite{CQB19,Ba130}.  We denote such rotational mode as mode I in present work and visualize it in Fig.~1(d). Considering that it is to some extent different from the precession of the ideal TW picture, we discuss in detail the difference in Sec. IIIC.

With the increase of ratio ${\cal J}_m/{\cal J}_s$, the rotational mode I emerges at  $I=12.5\hbar$ and $I=16.5\hbar$ as shown in Fig.~\ref{fig:Aplotb}, and  $I=12.5\hbar$  in Fig.~\ref{fig:Aplotc}. For higher spins, rotational mode II shown in Fig.~\ref{fig:picture}(e) appears. It is a planar tilted rotation in the $sm$ plane for both yrast and yrare band, and
the tilted directions become closer to the $m$ axis compared with the excited states of mode I. This is reflected by the larger $\varphi$ values of the maxima of $\mathcal{P}(\theta, \varphi)$.
In Ref.~\cite{Pd105}, the parameters ${\cal J}_{m}:{\cal J}_{s}:{\cal J}_{l}=1:0.63:0.22$ with $b=0.023$ were adopted, which are close to sets (B) and (C). The mode I
could be obtained for spin region $I\leq 29/2\hbar$ if such parameters are adopted,  which coincides with the conclusion of TW motion in Ref.~\cite{Pd105}.

For set (D) where ${\cal J}_m$ is significantly larger than ${\cal J}_s$ and close to hydrodynamical MOI, the results are illustrated in Fig.~\ref{fig:Aplotd}.
The $\varphi$ coordinates of the maxima of $\mathcal{P}(\theta, \varphi)$ for total angular momentum approach $0^{\circ}$ and  $180^{\circ}$ at $15.5\hbar$ of the yrast band.
For the excited state $16.5\hbar$ of the yrare band, the total angular momentum oscillates slightly with respect to the $m$ axis. It is indicated the core angular momentum $\textit{\textbf{R}}$ arises a wobble about the $m$ axis which is achieved by adding angular momentum component along the $s$ axis. This mode is called mode III and shown in Fig.~\ref{fig:picture}(f).
With the increasing spin, the distributions of the valence nucleon angular momenta are more diffuse.
Since the Coriolis force tends to realign $\textit{\textbf{j}}$ to the orientation of $\textit{\textbf{I}}$, the $m$ axis is preferred by valence nucleon angular momentum instead of $s$ axis at $I=19.5\hbar$ of the yrast band.
As the valence nucleon go back to the $s$ axis at $I=20.5\hbar$, the yrare band still maintains mode III.
Mode III is different from the LW mode shown in Fig.~\ref{fig:picture}(c), where the valence nucleon angular momentum steadily aligns along the $m$ axis.

The corresponding relationship of the azimuthal plots in Figs.~\ref{fig:Aplota}-\ref{fig:Aplotd} and the geometry of angular momentum  shown in Fig.~\ref{fig:picture} is summarized in Table~\ref{tab:pictureset}.
Based on the above discussion of the results adopted four sets of parameters,
we get the case which can be approximately represented as the TW motion when the ratio ${\cal J}_m/{\cal J}_s$ is closer to that of rigid-body MOI. The LW motion is approximately appear when the ratio is close to that of hydrodynamical MOI.
The variation of these angular momentum geometries is dependent on the values of  MOI, especially the ratio ${\cal J}_m/{\cal J}_s$. The mechanism of rotational bands in $^{105}$Pd might be associated with one of these modes or a mixture.
The present results also suggest that the
available data may not be sufficient to identify the TW mechanism, and further experimental explorations such as lifetime or
$g$-factor measurements are expected.

In Table~\ref{table:MOI}, we summarize the adopted MOI in the study of wobbling band based on PRM in the previous works, and the corresponding suggested wobbling modes. It can be seen that the calculations for $^{161,163,165,167}$Lu~\cite{Raduta2020PRC} and $^{187}$Au~\cite{Au187}, in which the adopted ${\cal J}_m/{\cal J}_s$ close to hydrodynamical MOI, have suggested the LW mode.
If the ratio is close to the rigid MOI or between the hydrodynamical and rigid MOI, then TW mode is in general suggested by PRM for $^{105}$Pd~\cite{Pd105}, $^{130}$Ba~\cite{Ba130}, $^{135}$Pr~\cite{Pr135,Frauendorf14,La133}, $^{163}$Lu~\cite{Frauendorf14}, and $^{183}$Au~\cite{Au183}.
Our conclusion from the study of $^{105}$Pd is not in conflict with the the previous calculated results and gives the relationship of  MOI and wobbling modes. However, the character of the nuclear MOI is a longstanding problem in the research of rotational band. Recently, both of the empirical MOI extracted experimentally~\cite{Allmond17} and the theoretical MOI calculated by the microscopic cranking model~\cite{Frauendorf182} suggested that the ratios of MOI follow the hydrodynamical model. If the hydrodynamical MOI is adopted, the TW mode seems to be difficult to occur from the above discussion.
The further understanding of MOI is a key point to clarify the debates on TW mode.

\subsection{Precession or tunneling}

\begin{figure}[ht!]
\centering
\subfigure[Transverse wobbling(TW)]{\includegraphics[width=5.3cm]{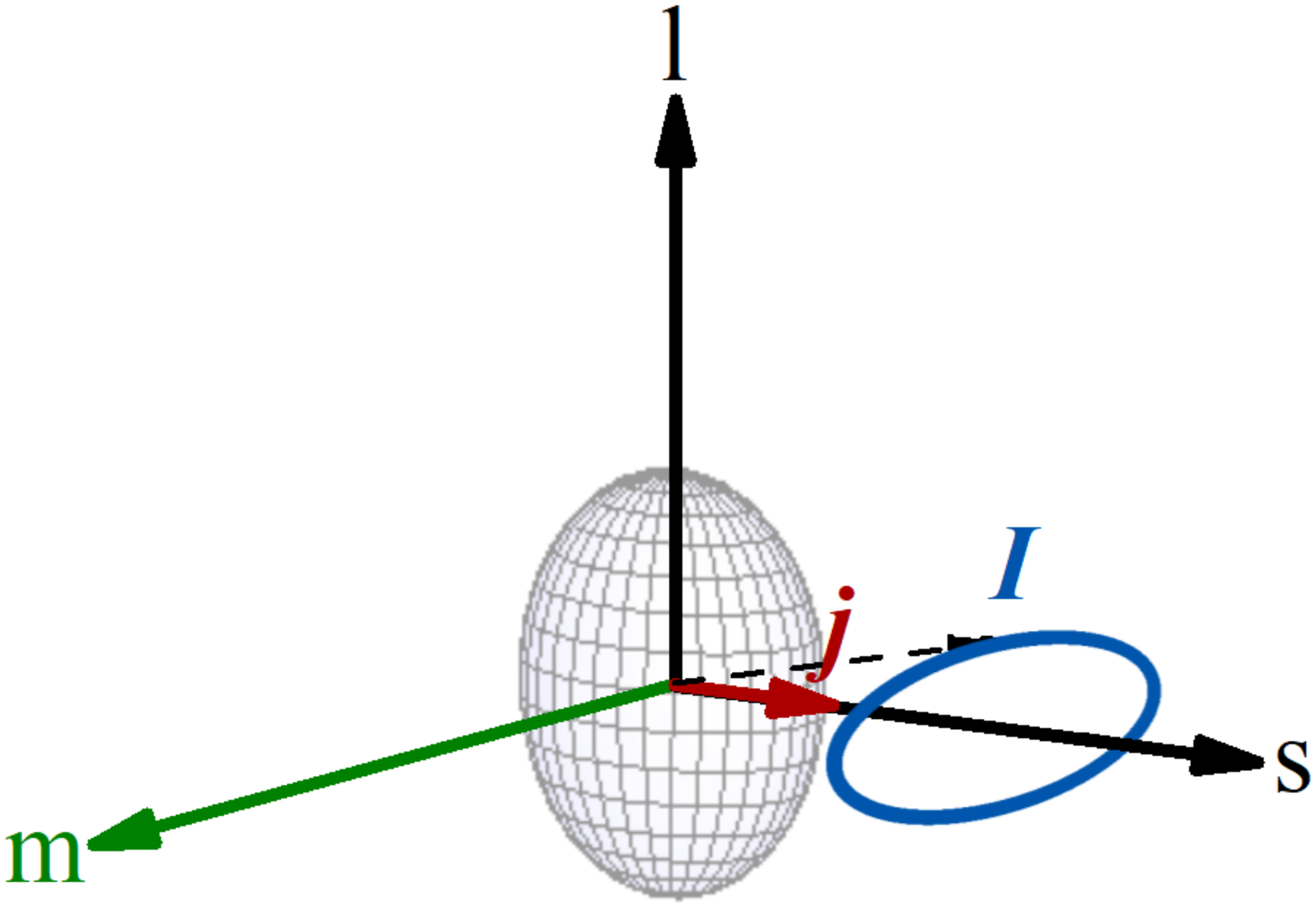}}
\subfigure[Tilted precession]{\includegraphics[width=5.3cm]{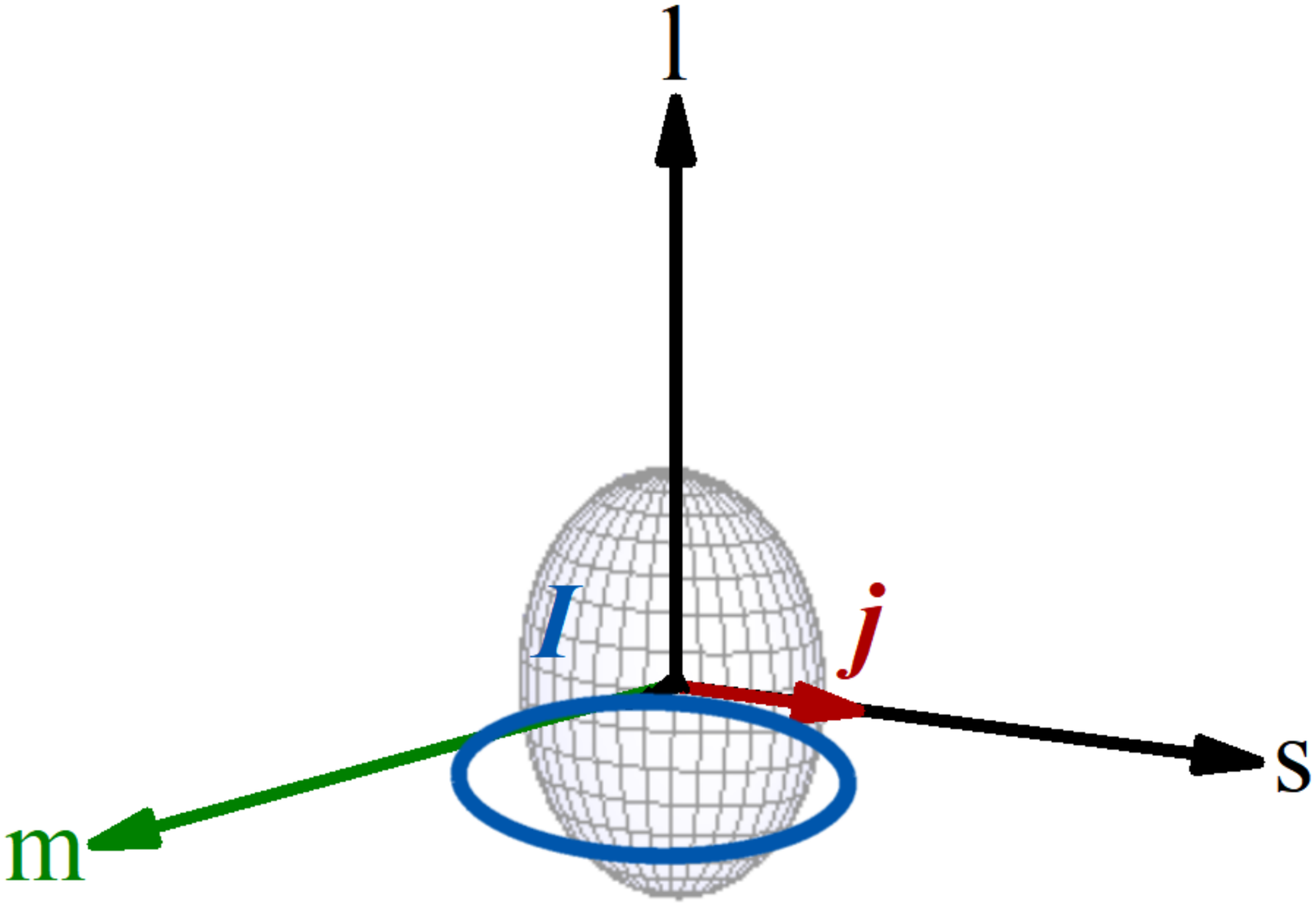}}
\subfigure[Longitudinal wobbling(LW)]{\includegraphics[width=5.3cm]{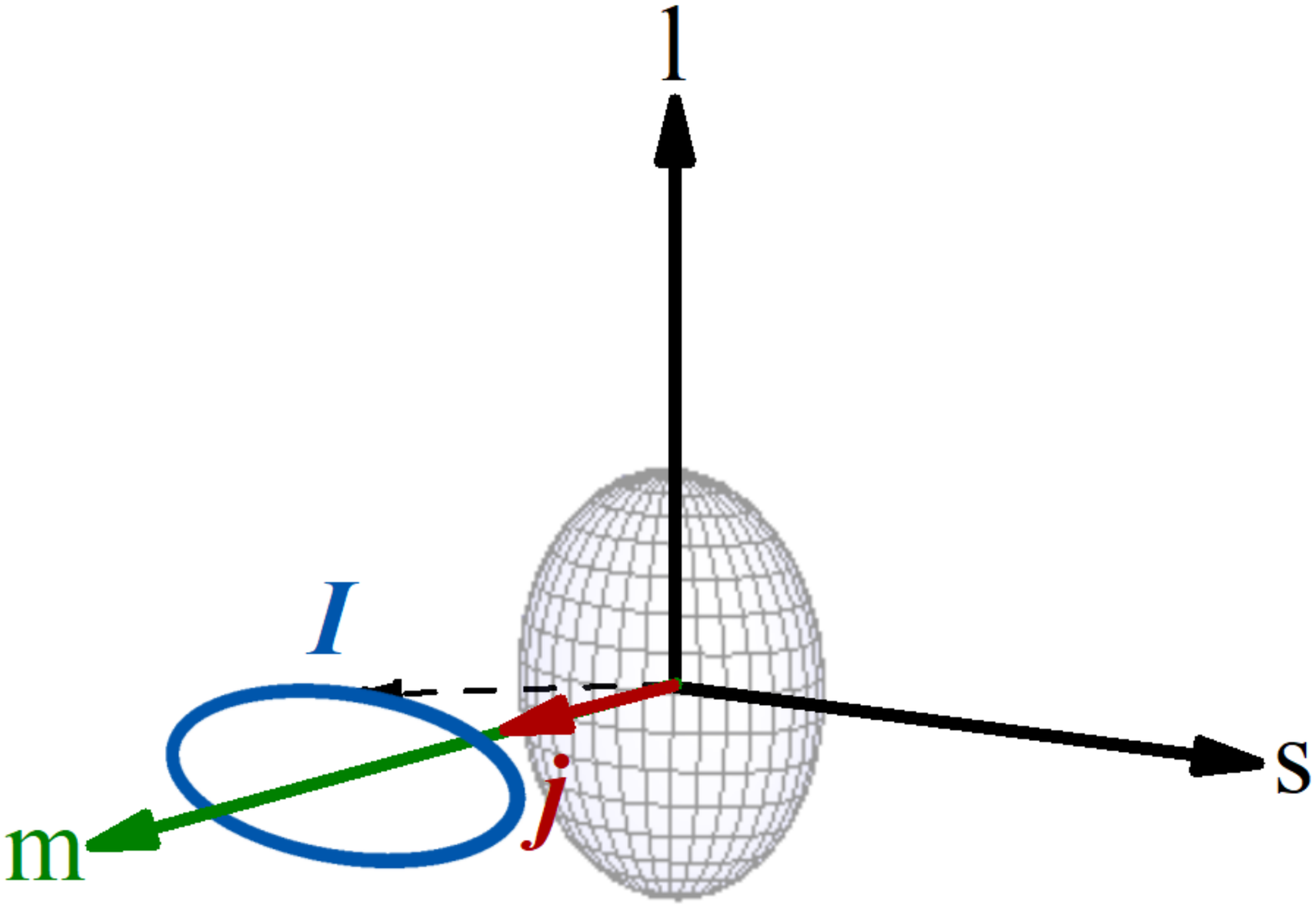}}\\
\subfigure[$^{105}$Pd. mode I]{\includegraphics[width=5.3cm]{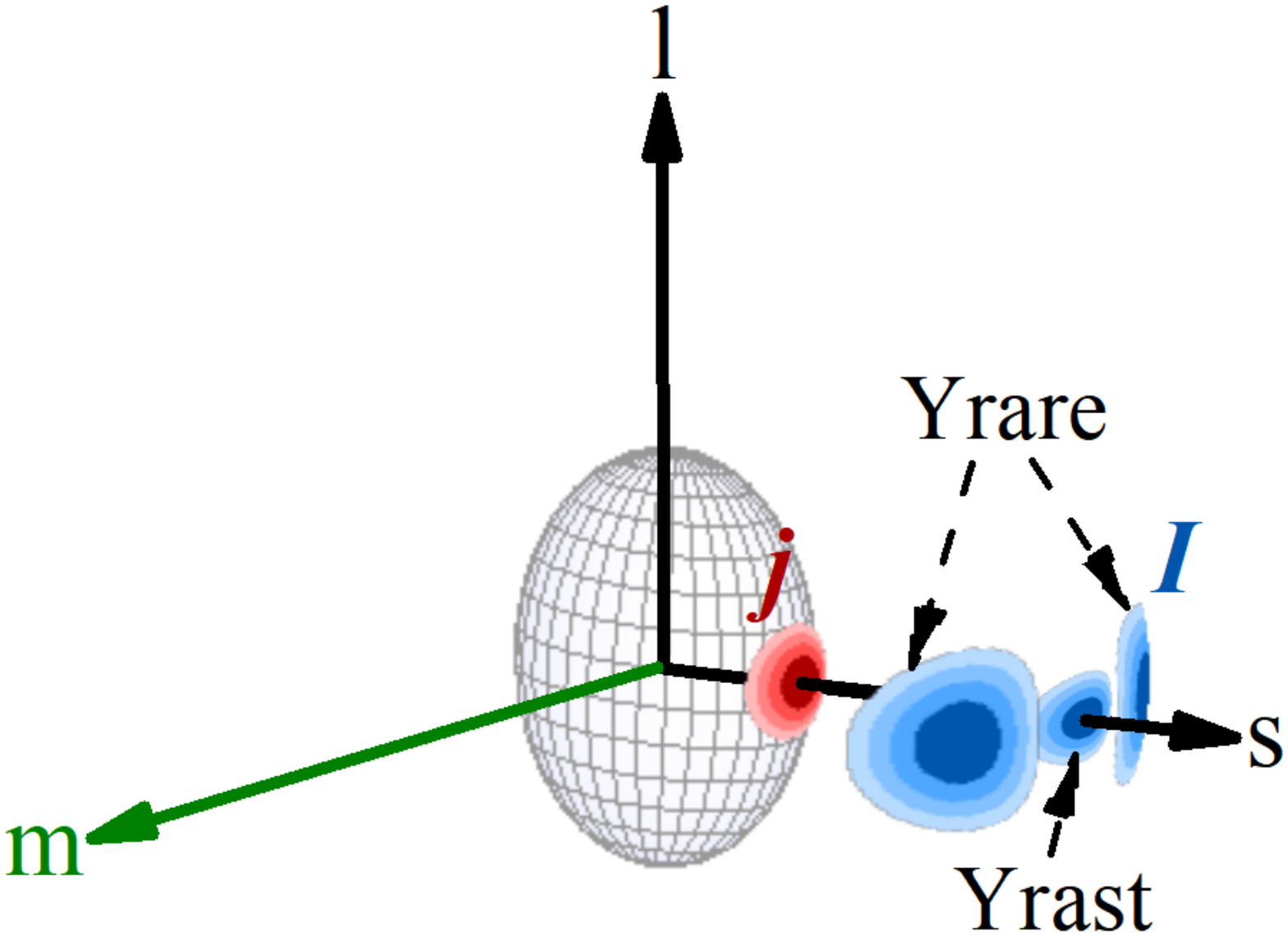}}
\subfigure[$^{105}$Pd. mode II]{\includegraphics[width=5.3cm]{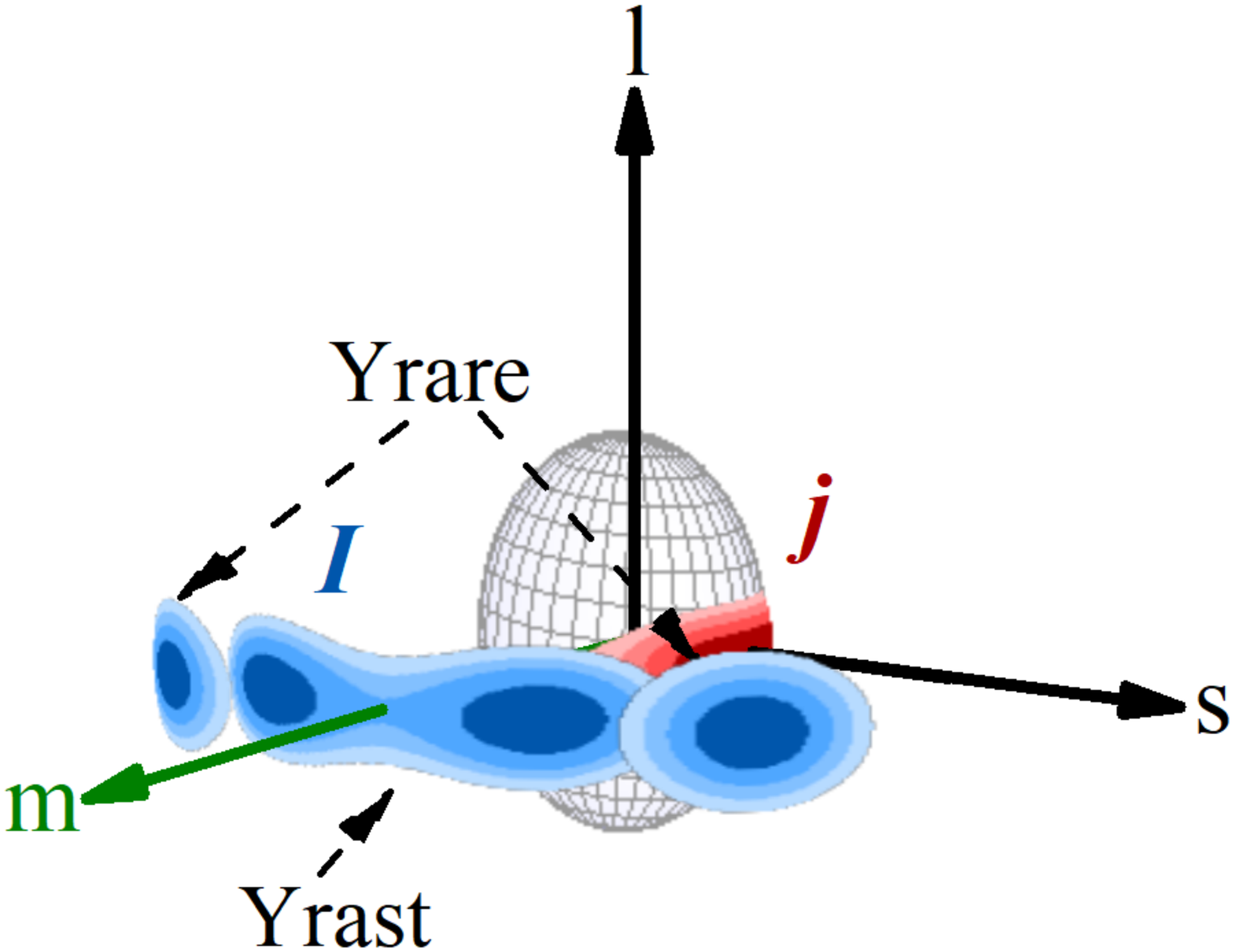}}
\subfigure[$^{105}$Pd. mode III]{\includegraphics[width=5.3cm]{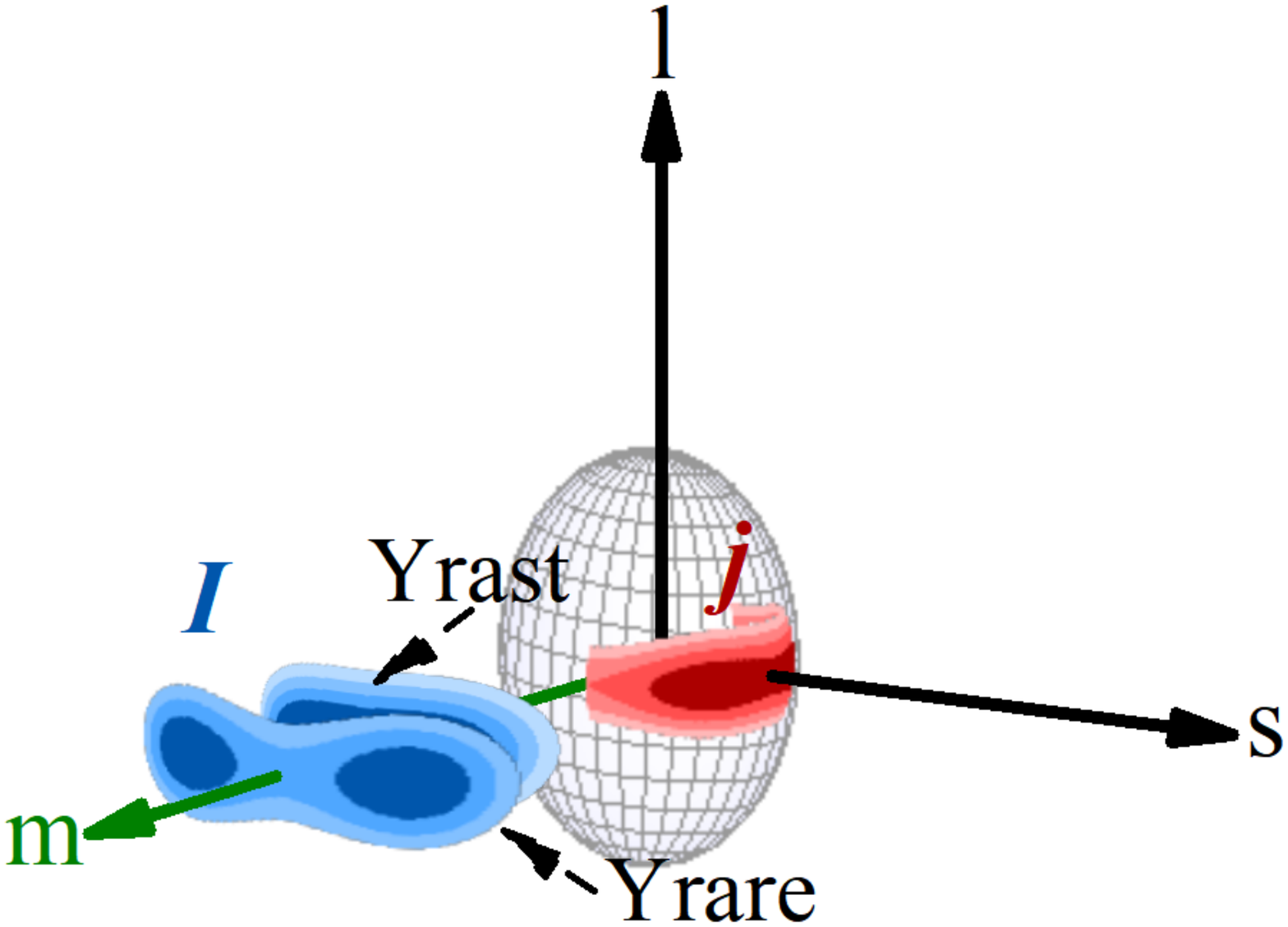}}
\caption{\label{fig:picture} Schematic diagram of angular momentum geometry for ideal transverse wobbling, longitudinal wobbling, and tilted precession, as well as the corresponding coupling scheme based on our calculations for $^{105}$Pd. The diagram of TW and LW refer to Ref.~\cite{Au187}.
The orientations of total and valence nucleon angular momentum are shown in blue and red, respectively.
Panels (d), (e), and (f) correspond to the quantum probability density distribution  at spin 15.5$\hbar$ (yrast)  and $16.5\hbar$ (yrare) calculated with  parameter sets (A), (C), and (D).
}
\end{figure}

\begin{figure}[ht!]
  \centering
\includegraphics[width=13cm]{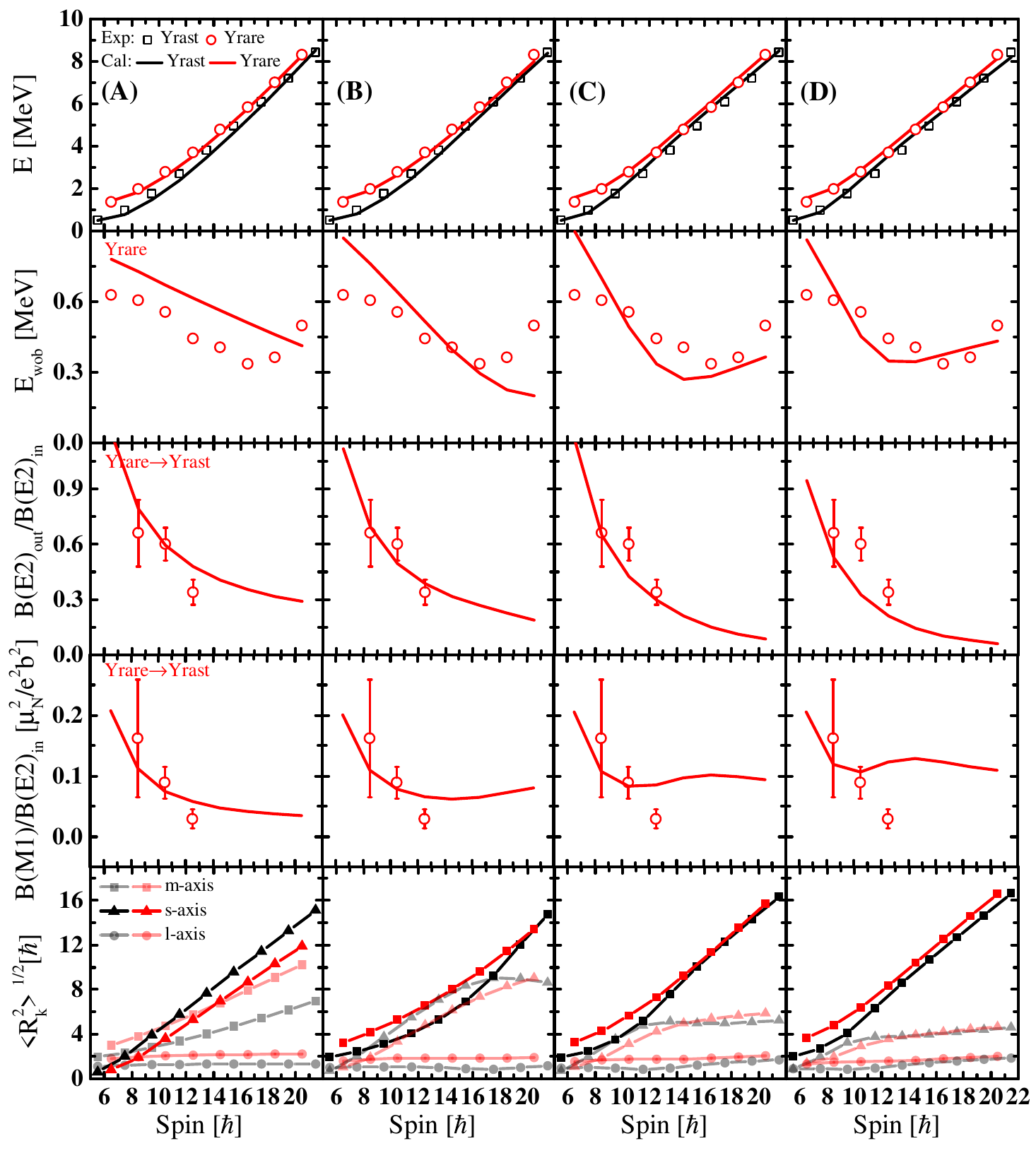}
\caption{\label{fig:reproduce}
The  energy spectra, wobbling energies, reduced transition probability ratio $B(E2)_{\textrm{out}}/B(E2)_{\textrm{in}}$, $B(M1)/B(E2)_{\textrm{in}}$, and root-mean-square values of the core angular momentum components as functions of the spin $I$ calculated by PRM (lines) in comparison to the experimental data of $^{105}$Pd~\cite{Pd105}(dots). The results with four parameter sets of MOI (A), (B), (C), and (D) in Table~\ref{tab:Pd105MOIset} are shown. The yrast and yrare band are respectively denoted by black and red.
}
\end{figure}

\begin{figure}[ht!]
  \centering
\includegraphics[width=7.5cm]{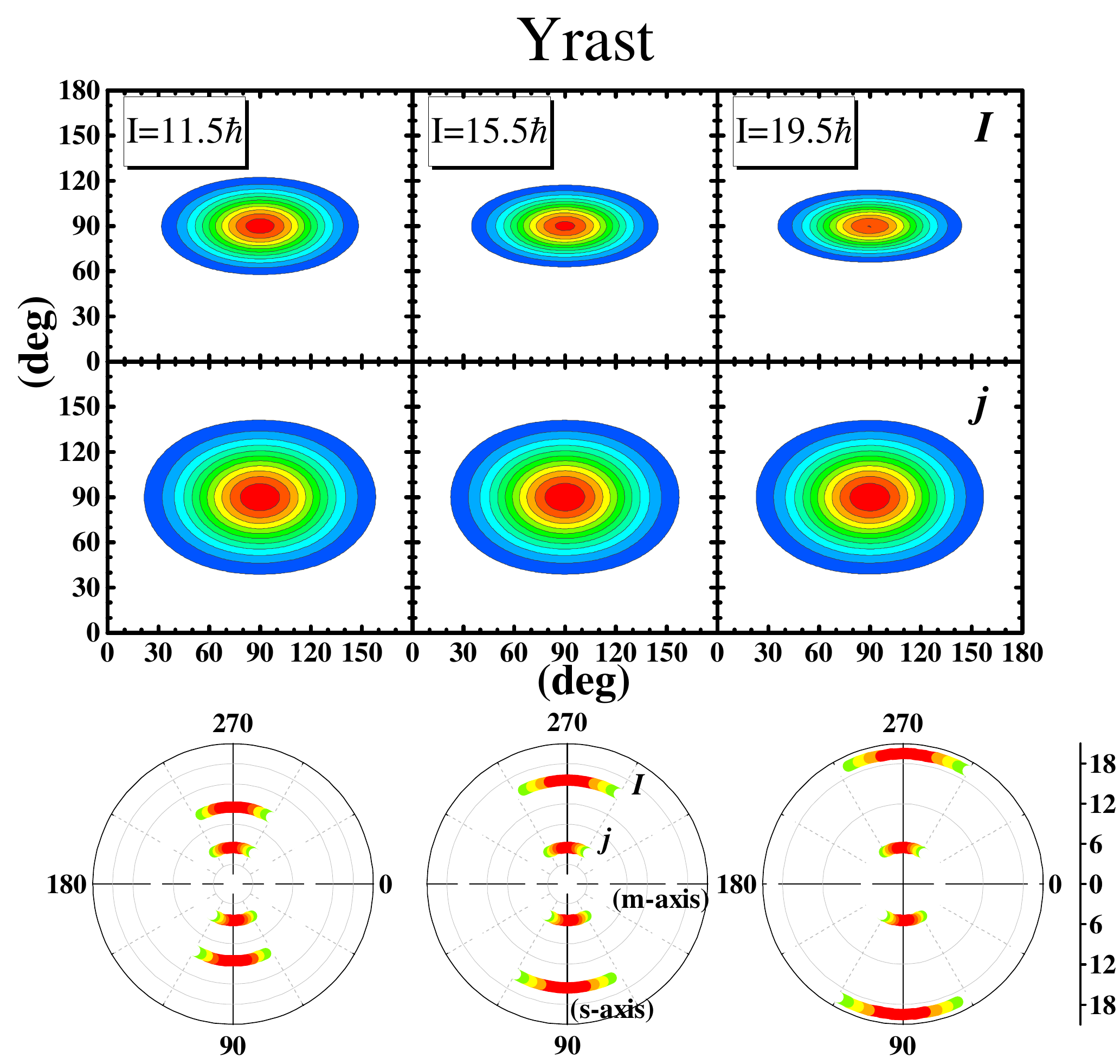}
\includegraphics[width=7.5cm]{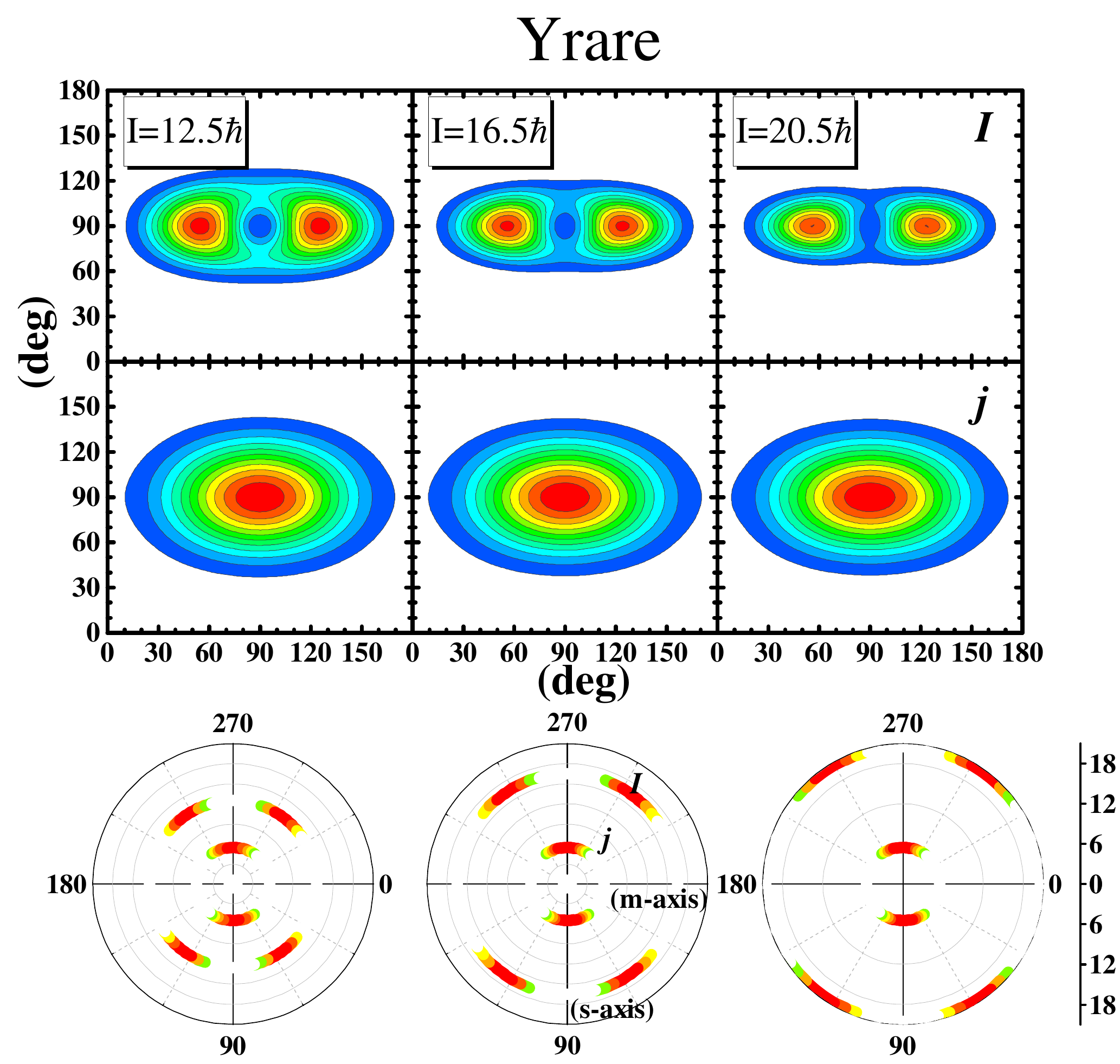}
\caption{\label{fig:Aplota}
Upper panels: The calculated probability distribution for the orientation of the angular momentum $\textit{\textbf{I}}$ and $\textit{\textbf{j}}$ on the ($\theta,\varphi$) plane for the yrast  and  yrare band in $^{105}$Pd. The parameter set (A) of MOI, $i.e.$, ${\cal J}_m:{\cal J}_s:{\cal J}_l=1:0.9:0.3$, is adopted. Red (blue) indicates maximal (minimal) probability.
Lower panels: The corresponding probability distribution of angular momenta in the $sm$ plane ($\theta=90^{\circ}$) with same color scheme of upper panels. The radial coordinate represents the magnitude of angular momentum ranging from 0 to 21$\hbar$, while the angle $\varphi$ from $0^{\circ}$ to $360^{\circ}$.}
\end{figure}

\begin{figure}[ht!]
  \centering
\includegraphics[width=7.5cm]{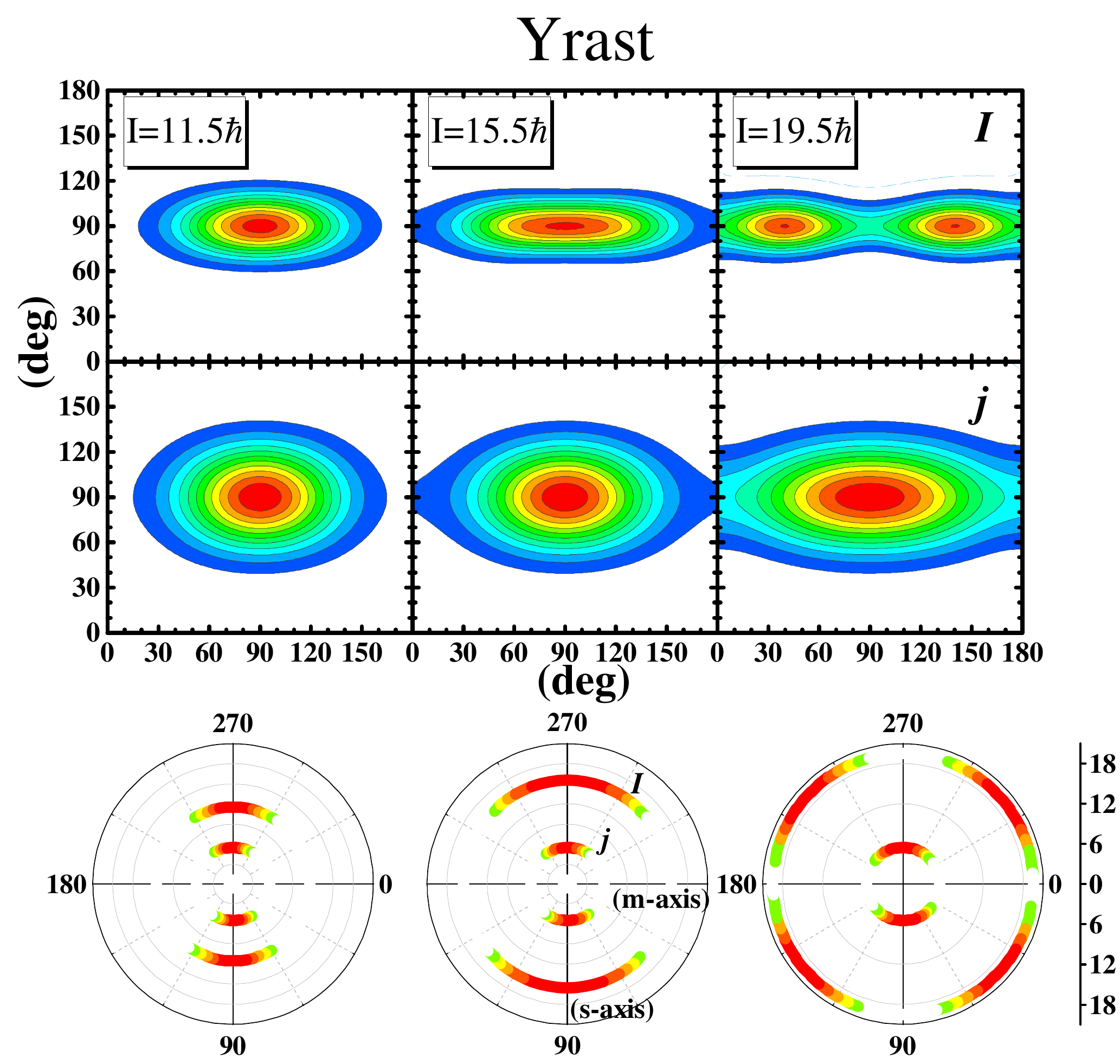}
\includegraphics[width=7.5cm]{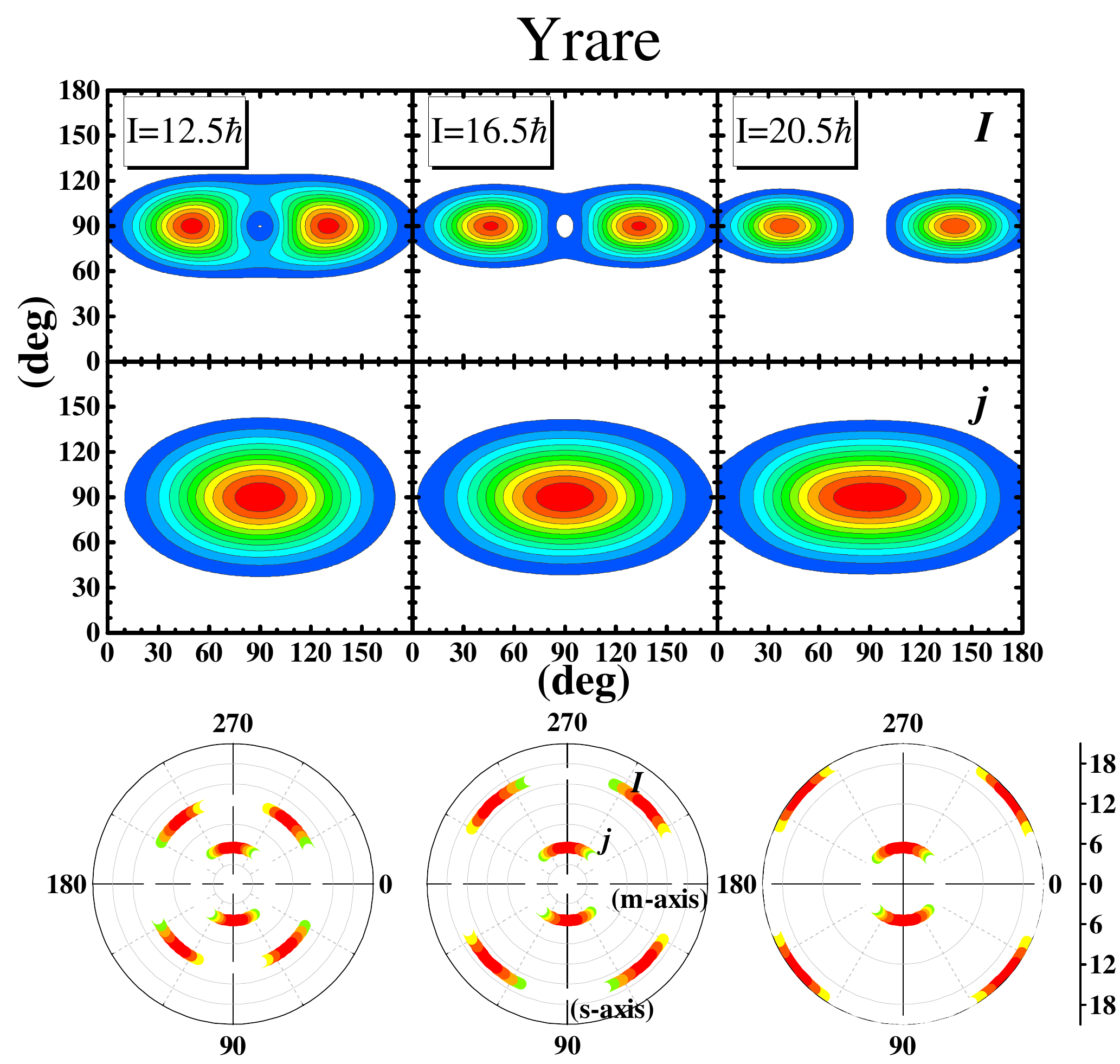}
\caption{\label{fig:Aplotb}
Same as Fig.~\ref{fig:Aplota} but for the results with MOI ${\cal J}_m:{\cal J}_s:{\cal J}_l=1:0.7:0.2$.
}
\end{figure}

\begin{figure}[ht!]
  \centering
\includegraphics[width=7.5cm]{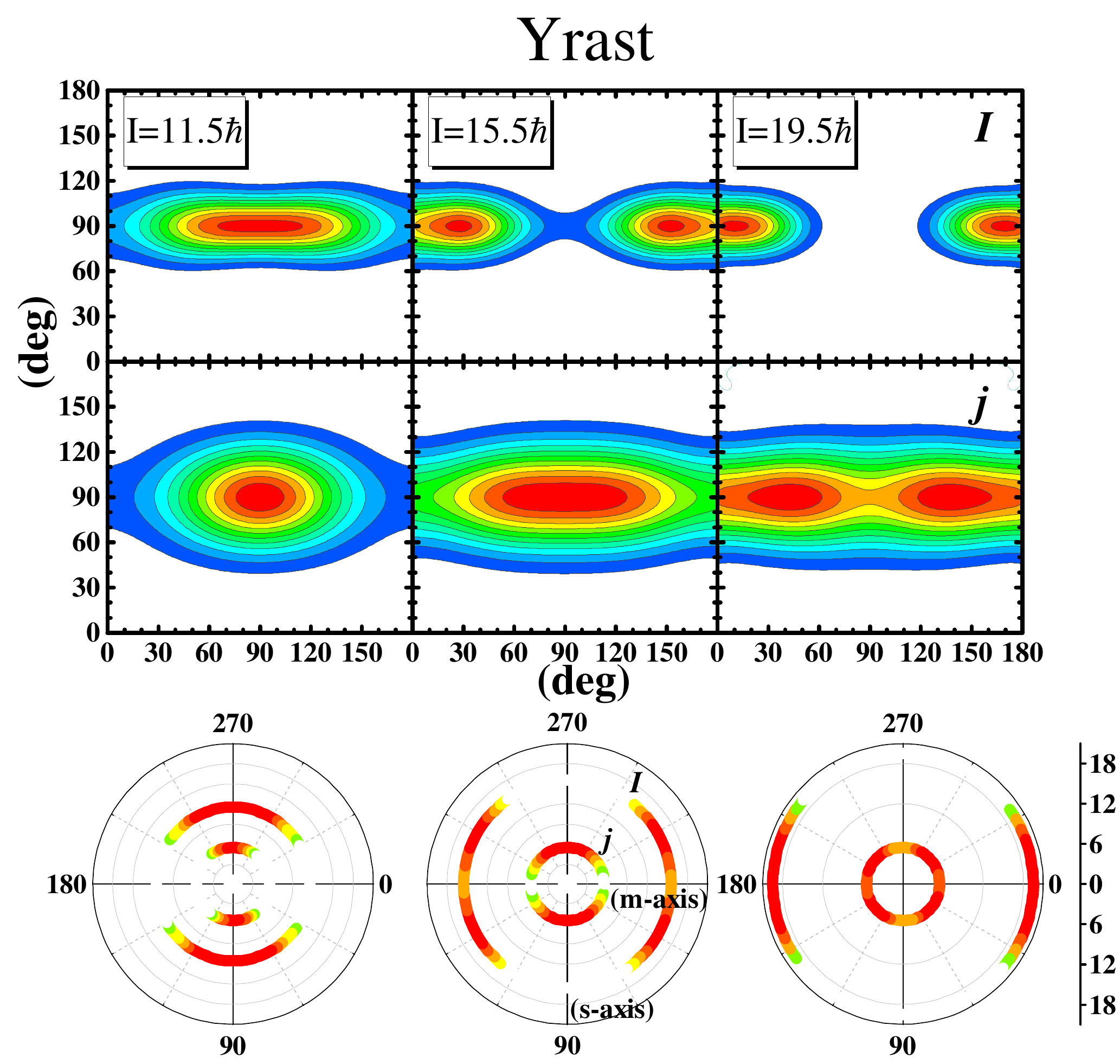}
\includegraphics[width=7.5cm]{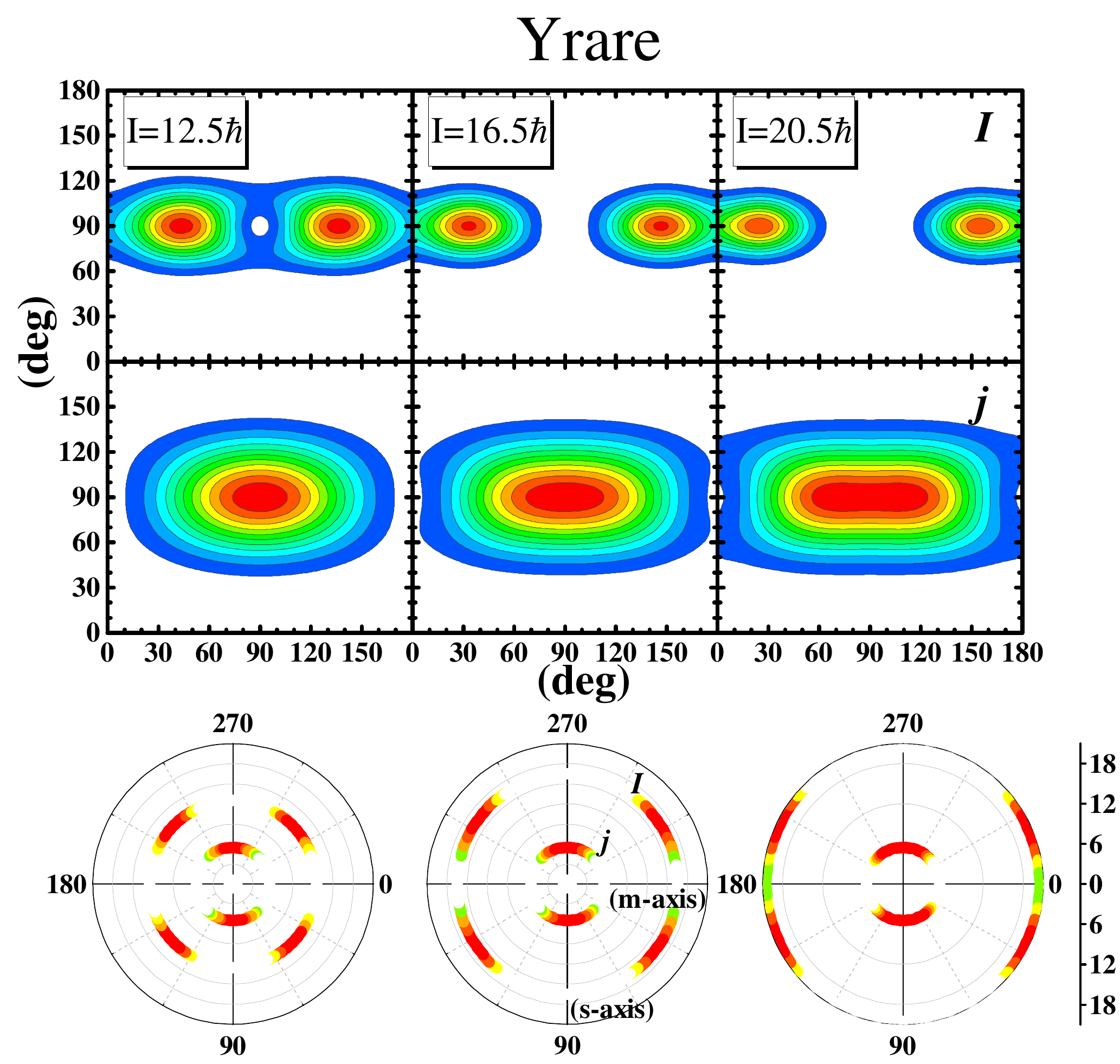}
\caption{\label{fig:Aplotc}
Same as Fig.~\ref{fig:Aplota} but for the results with MOI ${\cal J}_m:{\cal J}_s:{\cal J}_l=1:0.5:0.17$.
}
\end{figure}

\begin{figure}[ht!]
\centering
\includegraphics[width=7.5cm]{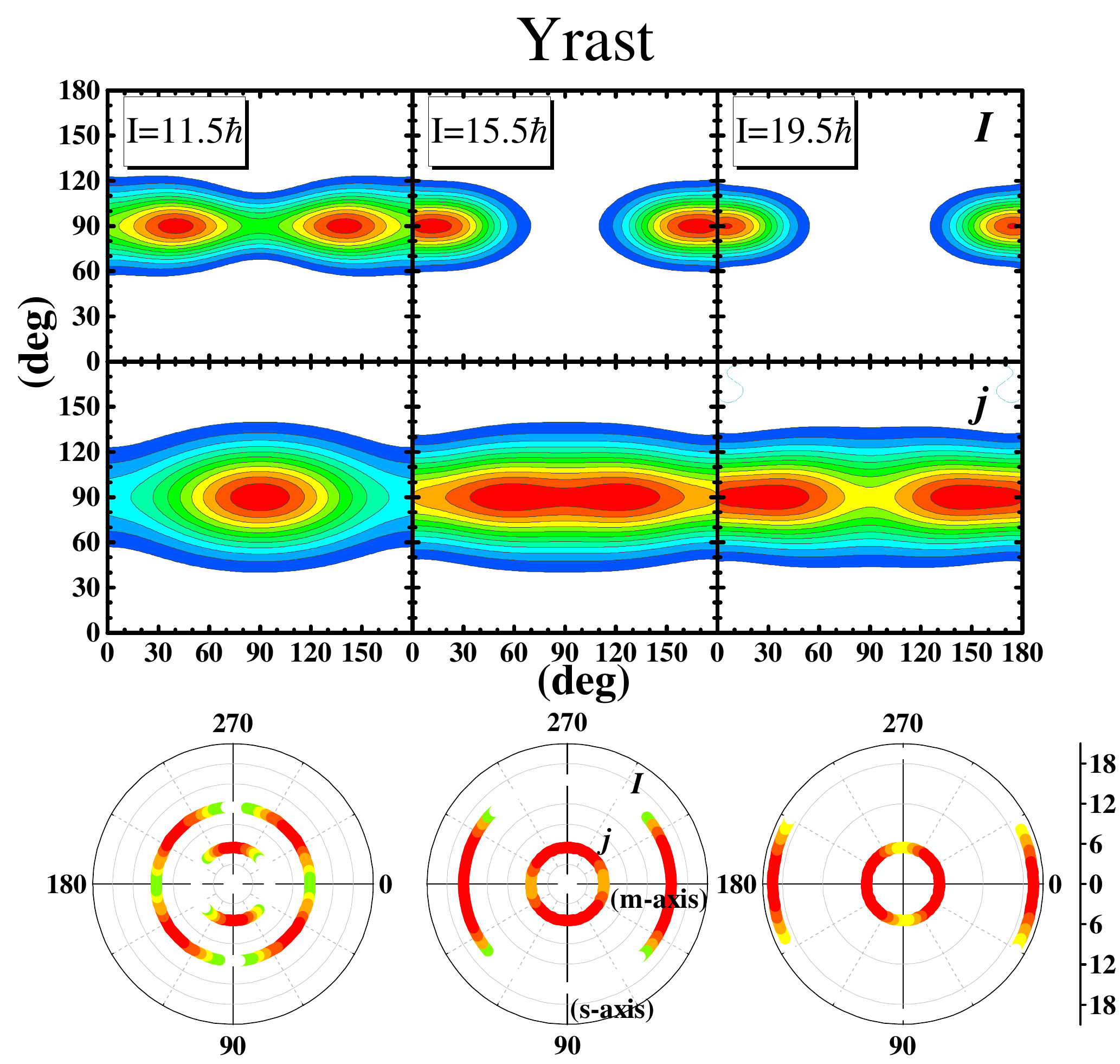}
\includegraphics[width=7.5cm]{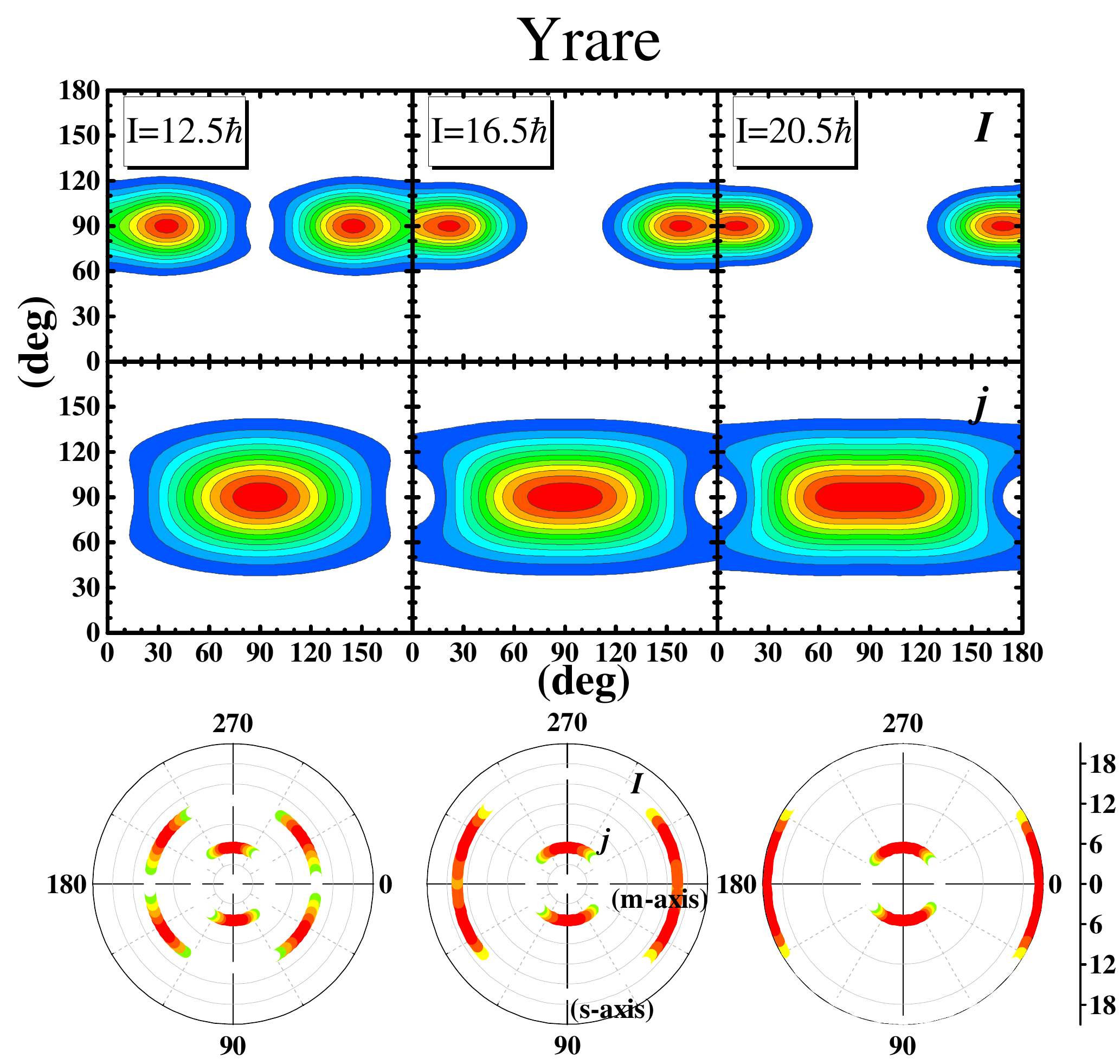}
\caption{\label{fig:Aplotd}
Same as Fig.~\ref{fig:Aplota} but for the results with  MOI ${\cal J}_m:{\cal J}_s:{\cal J}_l=1:0.3:0.08$.
}
\end{figure}

Precession of the nucleus means that  the  orientation of the total angular momentum is rotating about a principal axis.
Such precession has always been considered in the discussions about the TW, LW, or the tilted precession, as shown in Figs.~\ref{fig:picture}(a), \ref{fig:picture}(b), and \ref{fig:picture}(c). It should be noted the classical trajectories of  angular momentum for precession might change to tunneling in the quantum system.

\begin{figure}[ht!]
\centering
{\includegraphics[width=4cm]{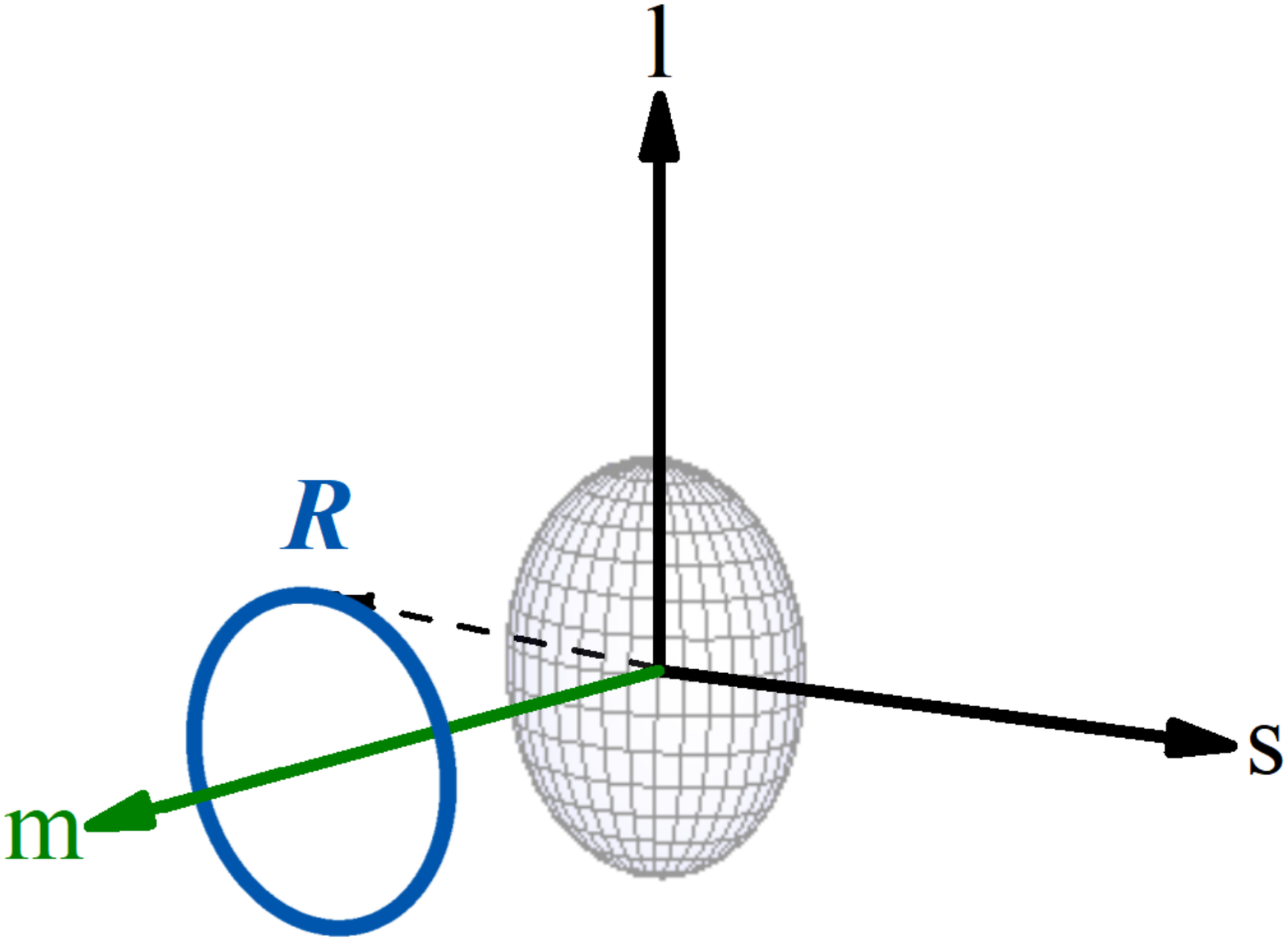}}
{\includegraphics[width=4cm]{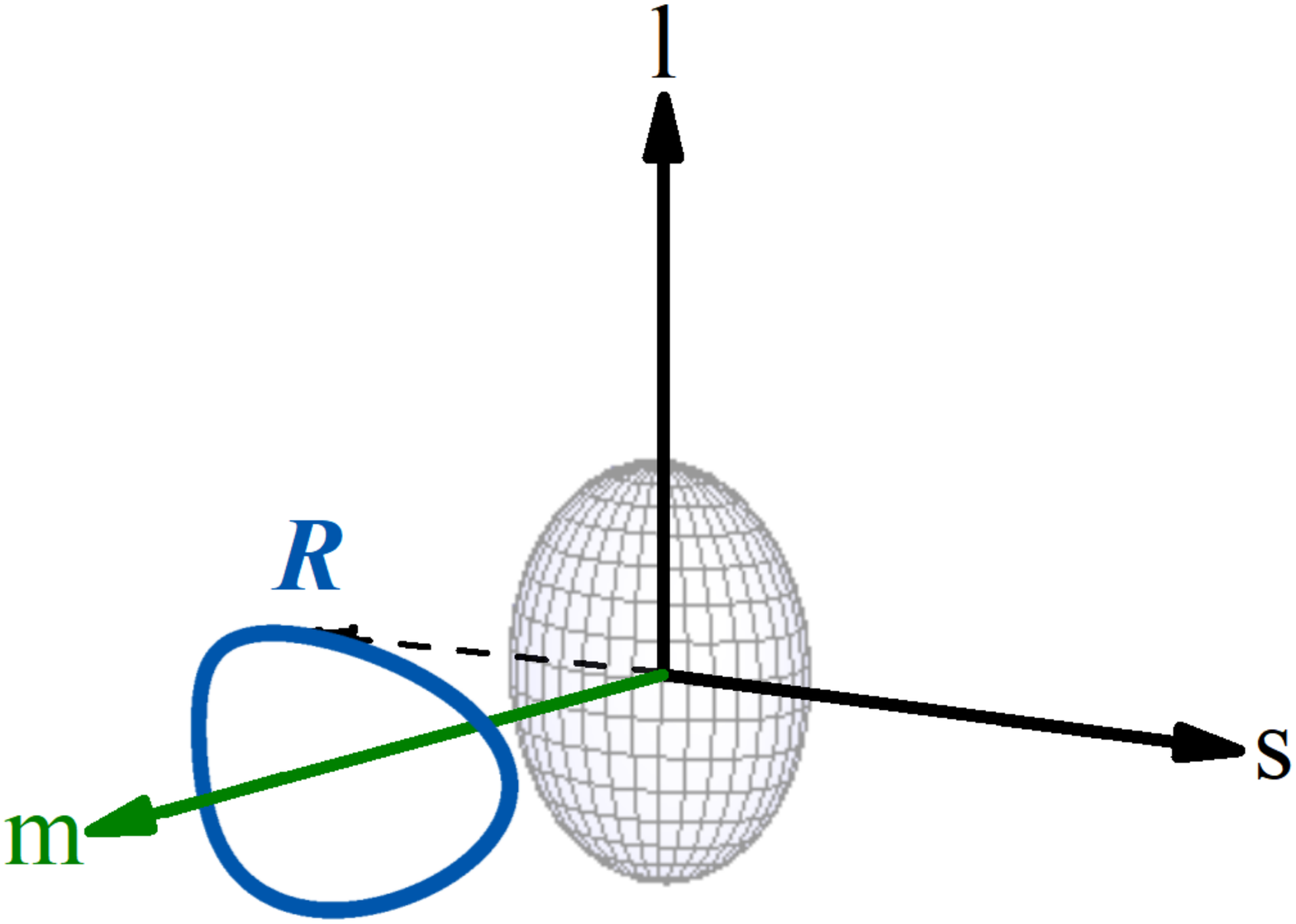}}
{\includegraphics[width=4cm]{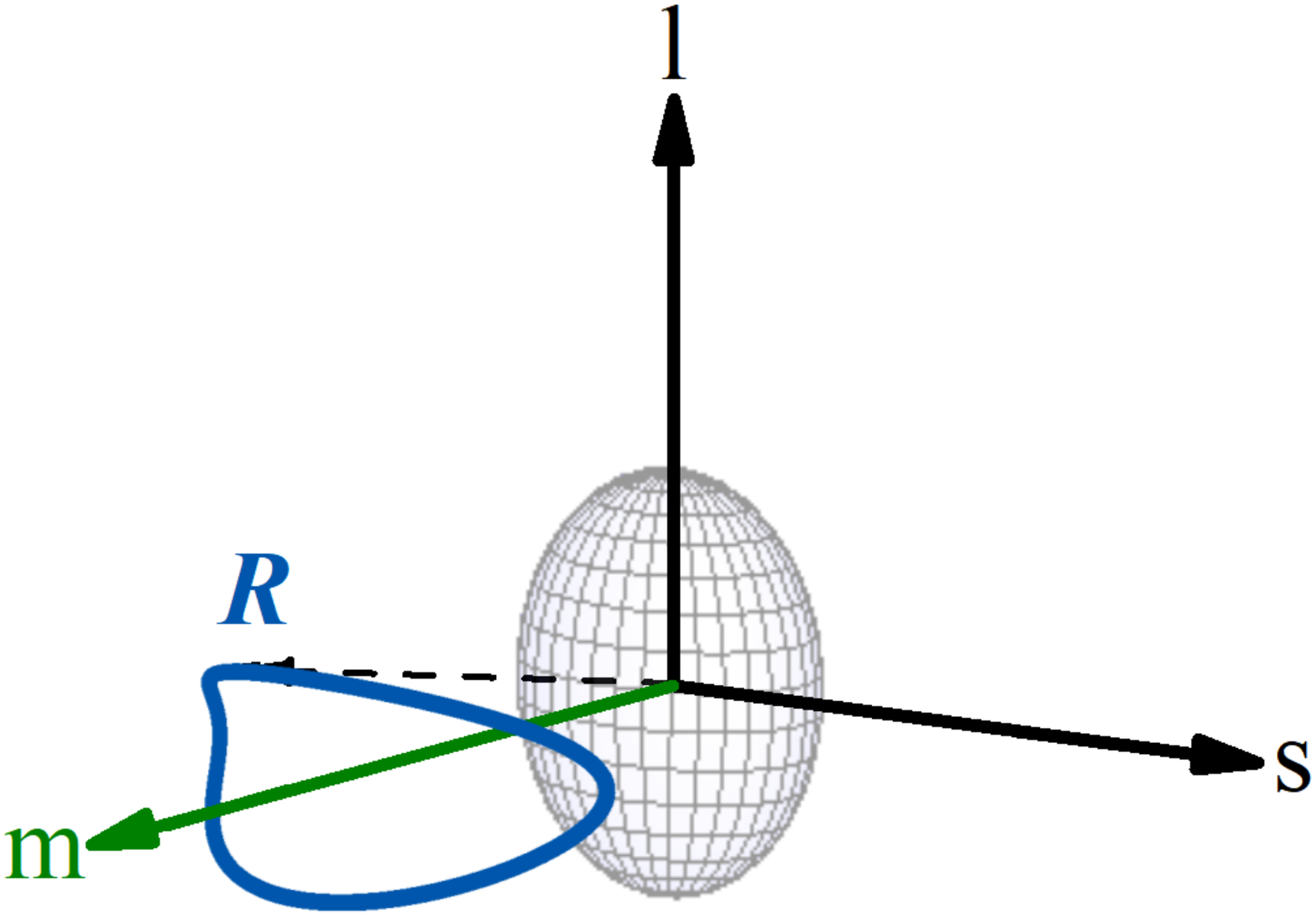}}
{\includegraphics[width=4cm]{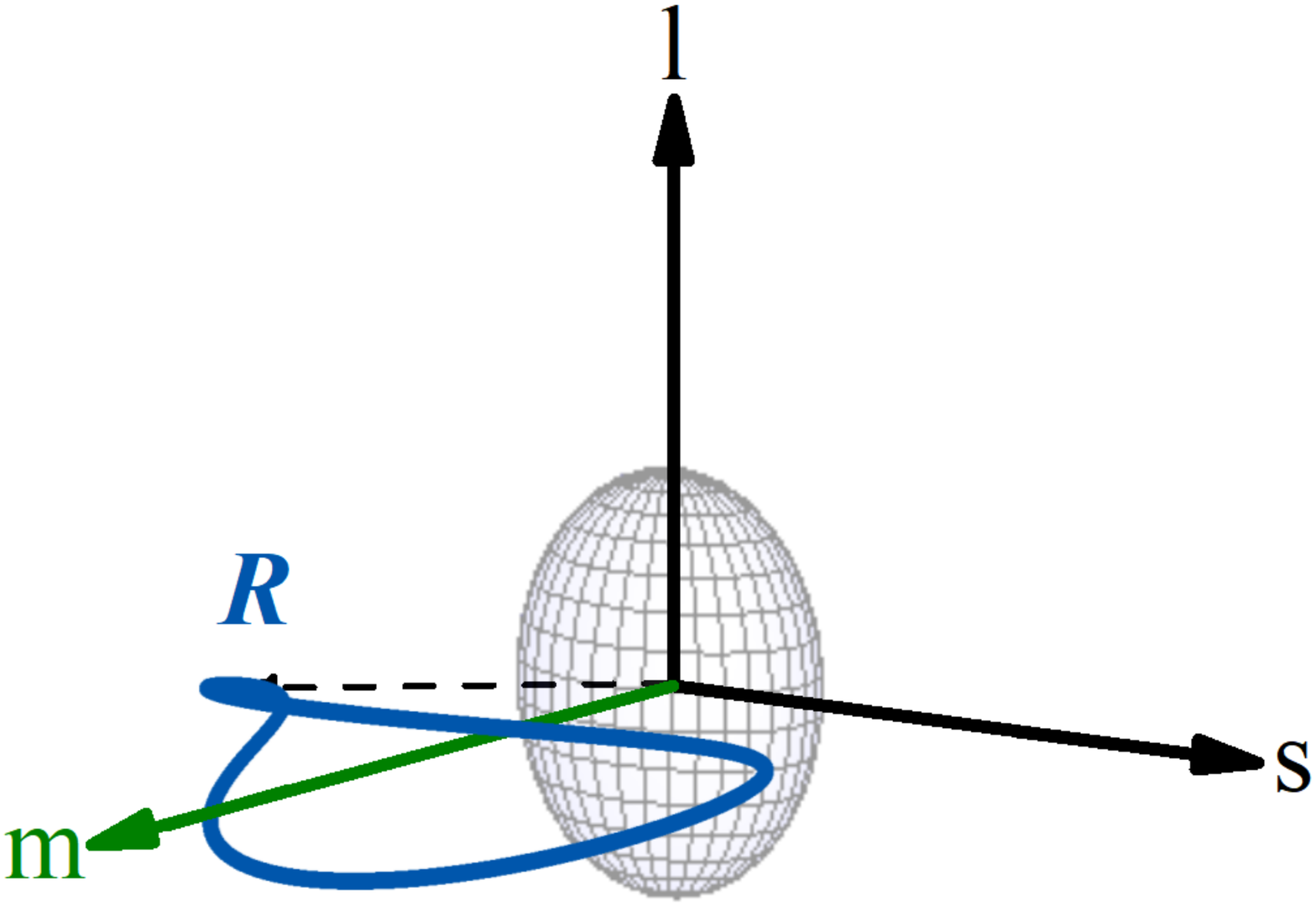}}\\
\subfigure[]{\includegraphics[width=4cm]{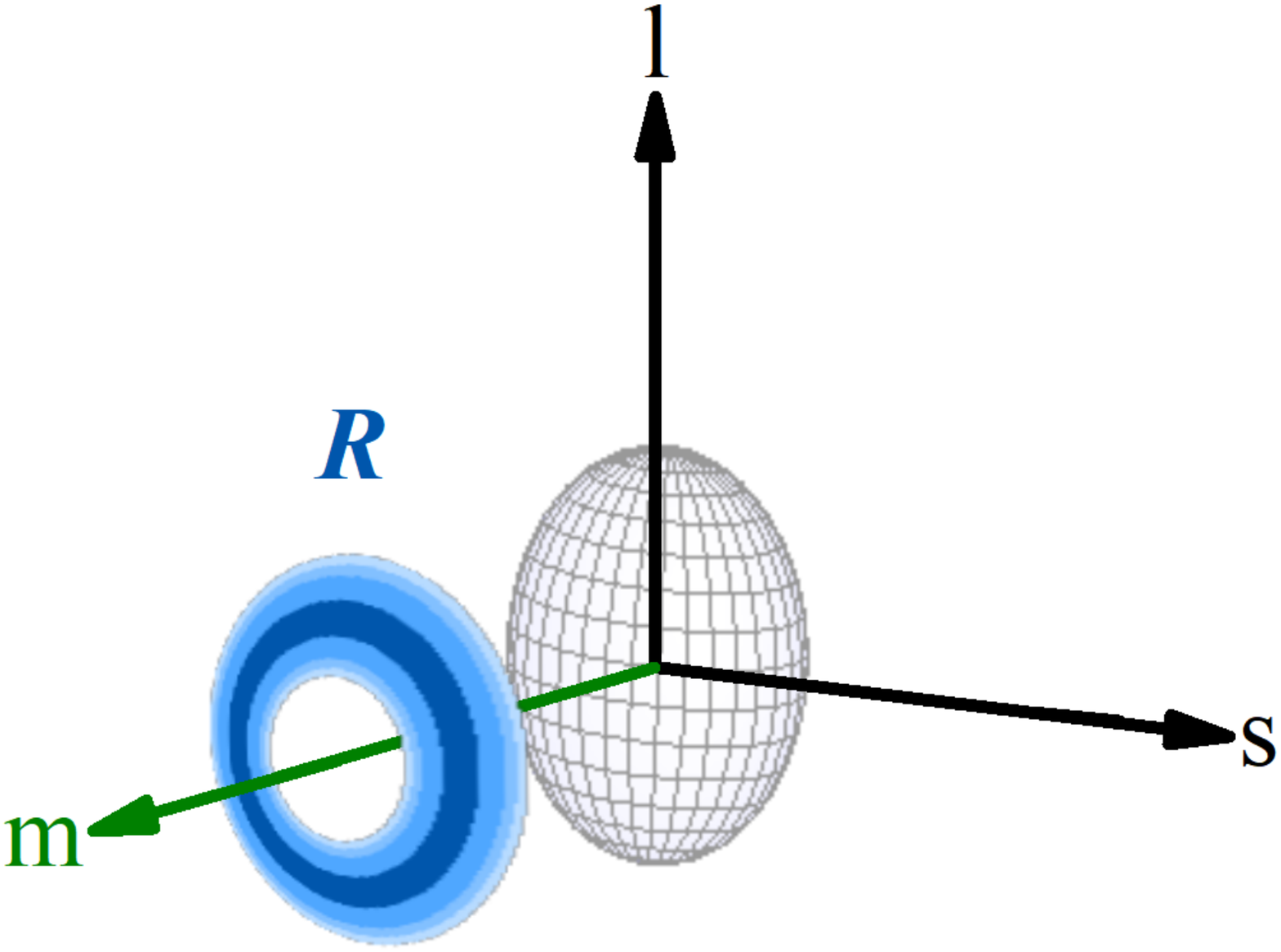}}
\subfigure[]{\includegraphics[width=4cm]{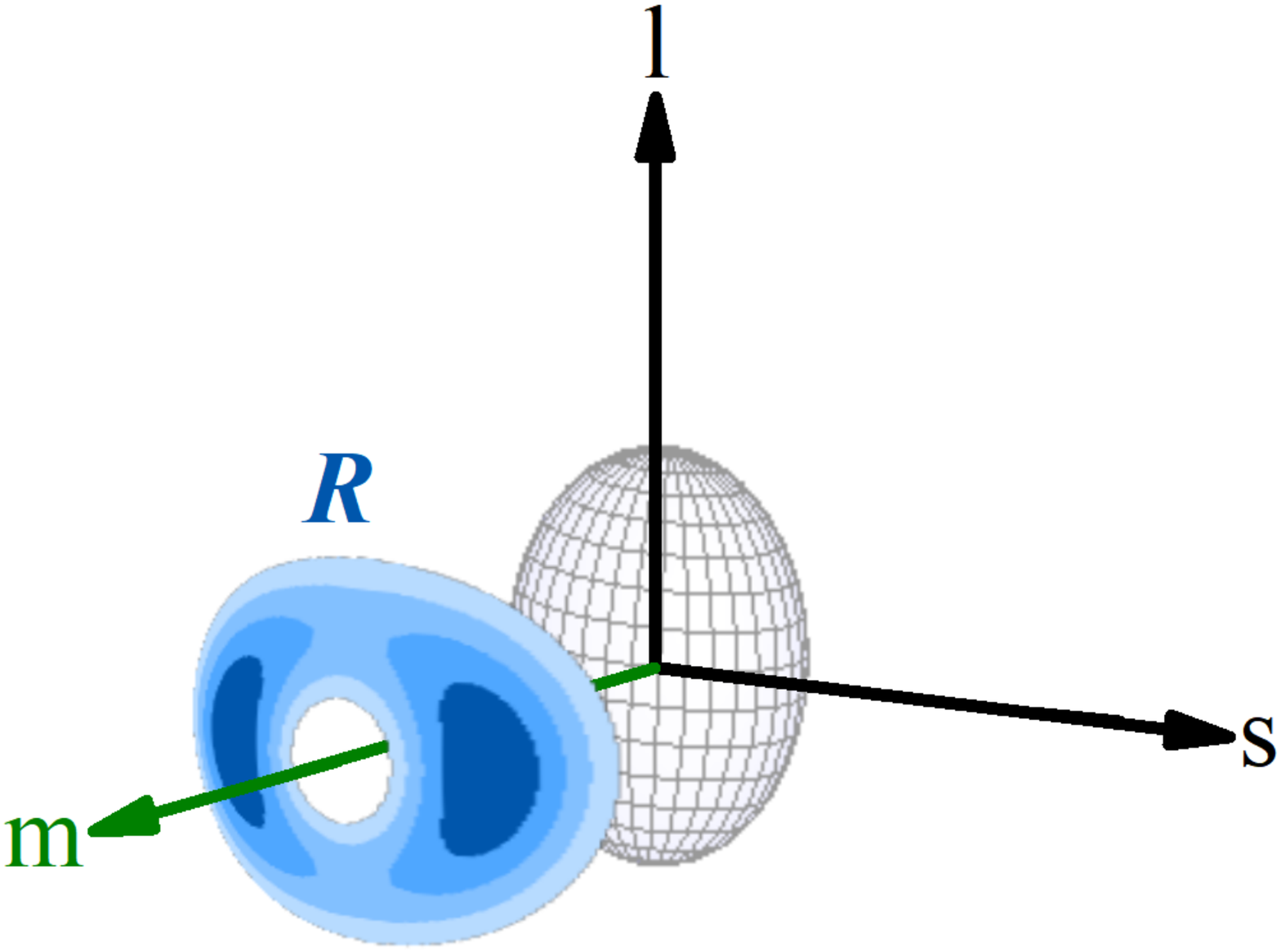}}
\subfigure[]{\includegraphics[width=4cm]{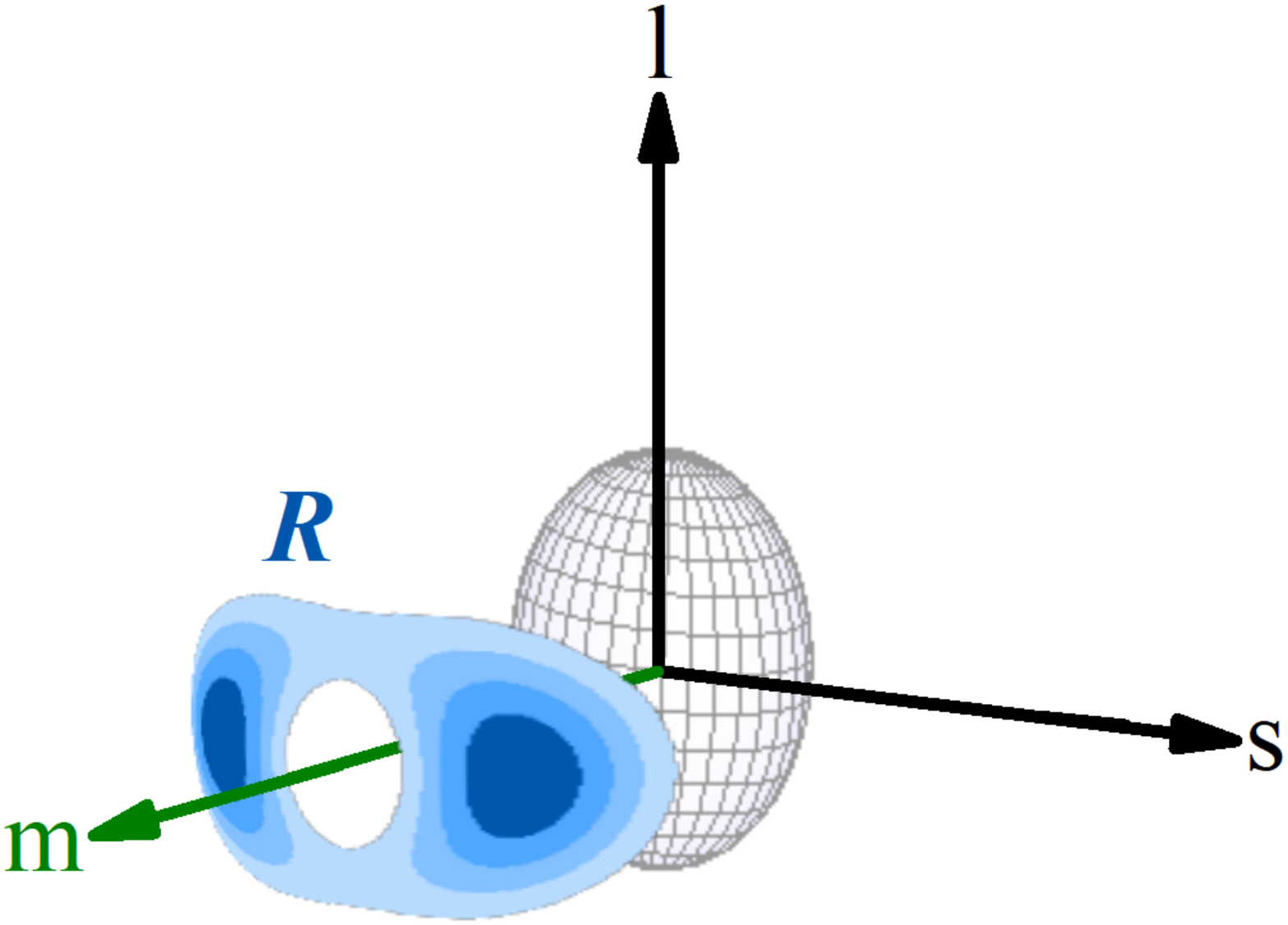}}
\subfigure[]{\includegraphics[width=4cm]{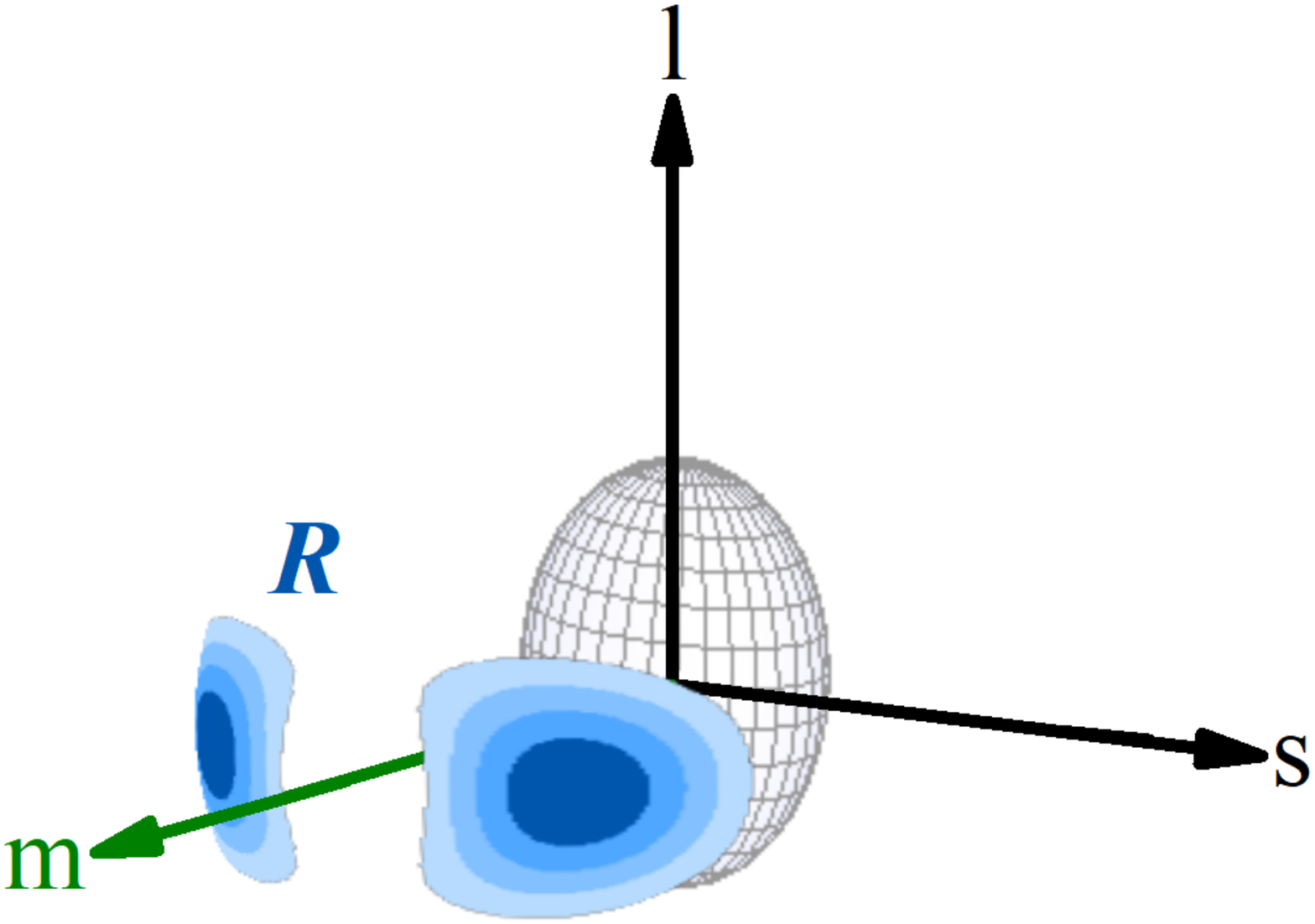}}\\
\caption{\label{fig:trajectory} The classical trajectory (upper panels) and quantum probability density distribution (lower panels) of angular momentum for a triaxial rotor at spin $13\hbar$  with differnent MOI.  The ratios ${\cal J}_{m}/{\cal J}_{s}/{\cal J}_{l}$ are respectively $1/0.25/0.25$ (a), $1/0.33/0.18$ (b), $1/0.43/0.12$ (c), and $1/0.54/0.07$ (d). The obtained quantum energy $E=1.48, 1.48, 1.49, 1.51$MeV are  input to get the classical trajectory. }
\end{figure}

To address
this issue, we provide the classical trajectories and the quantum probability density distribution of angular momentum in Fig.~\ref{fig:trajectory}. Here, we plot the case of purely collective rotor to avoid the ambiguity due to coupling of a particle to a rotor core.
The classical orbits of angular momentum are obtained from  the intersection of the angular momentum sphere
$R_{m}^{2}+R_{s}^{2}+R_{l}^{2}=R(R+1)$
with the energy ellipsoid
$R_{m}^{2}/{\cal J}_{m}+R_{s}^{2}/{\cal J}_{s}+R_{l}^{2}/{\cal J}_{l}=2E$~\cite{Landau,Lawrie20,Frauendorf14}.
For the quantum probability density distribution of angular momentum, it is obtained by the quantum triaxial rotor model~\cite{Bohr75} with  the probability distribution calculated by Eq.~(\ref{APlot}).
In Figs.~\ref{fig:trajectory}(a), \ref{fig:trajectory}(b), \ref{fig:trajectory}(c) and \ref{fig:trajectory}(d), the ratios of MOI ${\cal J}_{m}/{\cal J}_{s}/{\cal J}_{l}$ are adopted respectively as $1/0.25/0.25$, $1/0.33/0.18$, $1/0.43/0.12$, and $1/0.54/0.07$. These ratios are corresponding to the hydrodynamical MOI with $\gamma=30^{\circ},25^{\circ},20^{\circ},15^{\circ}$.
The obtained energy $E$ calculated by quantum triaxial rotor model for spin $13\hbar$ are respectively $1.48, 1.48, 1.49$, and $1.51$ MeV, which are input to get the classical trajectories.

The classical trajectories  and quantum probability density distribution  of the angular momentum are circles only in the case of ${\cal J}_{s}={\cal J}_{l}$, as shown in Fig.~\ref{fig:trajectory}(a).
In general, the classical trajectory are not circles, as shown in Figs.~\ref{fig:trajectory}(b), \ref{fig:trajectory}(c), and \ref{fig:trajectory}(d). Correspondingly,
the maximum probabilities of the angular momentum locate at two symmetrically points about the $m$ axis in the quantum case. A tunneling should occur if the barrier between them is high enough.
Thus,  precession and tunneling are two aspects of the quantum wobbling motion.
Modes I, II, and III in our calculation for $^{105}$Pd, as shown in Figs.~\ref{fig:picture}(d), \ref{fig:picture}(e), \ref{fig:picture}(f), exhibit a tunneling between two symmetrically
orientations of angular momentum for the yrare band.
It seems that the tunneling between two orientations of angular momentum is preferable to constitute the TW and LW.

\section{Summary}

We have reinvestigated the reported  TW band in an odd-mass nucleus $^{105}$Pd based on the triaxial particle rotor model.
The experimental data of wobbling band could be
reproduced by the calculated results adopting different parameter sets of MOI. By analyzing the azimuthal plot of the angular momentum of nucleus and valence nucleon, different types of rotational modes are shown.
It is confirmed that these modes are sensitive to the ratio between the MOI at the $m$ and $s$ axis, namely  ${\cal J}_{m}/{\cal J}_{s}$.
The TW mode appears approximately when the ratio agrees with the rigid-body MOI, $i.e.$, the total angular momentum   wobbles around the $s$ axis.
When the ratio agrees with the hydrodynamical MOI, a wobbler around the $m$ axis occurs. When the ratio is between the above two, the planar tilted rotation occurs for both of the yrast and yrare band.
The mechanism of rotational bands in $^{105}$Pd might be one of these three modes. In addition, it is exhibited that
precession and tunneling are two aspects of the quantum wobbling motion. The tunneling aspect dominates
in the present yrare states of $^{105}$Pd.
The further understanding of nuclear MOI in theory is necessary to clarify the debates on wobbling motion, and further experimental explorations to identify the TW mechanism are also expected.

\begin{acknowledgments}
This work is partly supported by the Major Program of Natural Science Foundation of
Shandong Province (No.~ZR2020ZD30), the Outstanding Youth Fund of Natural Science Foundation
of Shandong Province (ZR2020YQ07), the National Natural Science Foundation of China
(No.~11675094, No.~12075137, No. 12075138), and the Shandong Natural Science Foundation (No.~ZR2020QA084).
\end{acknowledgments}

\clearpage

\renewcommand{\thefootnote}{\alph{footnote}}
\begin{longtable}{ccccclc}
\caption{The adopted parameter sets of MOI ${\cal J}_{k}=a_{k}\sqrt{1+bI(I+1)}$ in the calculation.}
\label{tab:Pd105MOIset} \\
\hline
\hline
Parameter Set~~ & ~~${a}_m$ ~~& ~~${a}_s$ ~~&~~ ${a}_l$ ~~ & b &~~~~${\cal J}_m:{\cal J}_s:{\cal J}_l$ & ~~~Behaviour
of ${\cal J}_m:{\cal J}_s$~~\\
\hline
(A) & 6.0 & 5.4 & 1.8 & ~ 0.016  &~~~ $1:0.9:0.3$ & Rigid body \\
(B) & 6.0 & 4.2 & 1.2 & ~ 0.023  &~~~ $1:0.7:0.2$  & Inbetween\\
(C) & 6.0 & 3.0 & 1.0 & ~ 0.026 &~~~ $1:0.5:0.17$ & Inbetween \\
(D) & 12.0 & 3.6 & 1.0 &~  0.008 &~~~ $1:0.3:0.08$ & Hydrodynamical\\
 \hline
\end{longtable}

\begin{longtable}{ccccccc}
\caption{The corresponding relationship of the azimuthal plots in Figs.~\ref{fig:Aplota}-\ref{fig:Aplotd} and the geometry of angular momentum  shown in Fig.~\ref{fig:picture}.}
\setlength{\tabcolsep}{7mm}
\label{tab:pictureset} \\
\hline
\hline
\multicolumn{2}{c}{Spin}  & \multicolumn{4}{c}{Corresponding rotational modes}     \\
    \hline
Yrast & Yrare & Parameter (A) & ~~Parameter (B)~ & Parameter (C)  & Parameter  (D) \\
    \hline
    $I=11.5\hbar$ & ~~~$I=12.5\hbar$ ~~~& Mode I & Mode I & Mode I &  Mode II \\
    $I=15.5\hbar$ & $I=16.5\hbar$ & Mode I & Mode I & Mode II &  Mode III \\
    \multirow{2}{*}{$I=19.5\hbar$} & \multirow{2}{*}{$I=20.5\hbar$} & \multirow{2}{*}{Mode I} & \multicolumn{1}{c}{\multirow{2}{*}{Mode II}} & \multirow{2}{*} {Mode III} & LW (yrast) \\
                       &                    &                    & \multicolumn{1}{c}{}                   &  \multicolumn{1}{c}{}  &  Mode III (yrare) \\
\hline
\end{longtable}

\begin{table}
\caption{The adopted MOI in the study of wobbling band based on PRM in the previous works, together with the corresponding suggested wobbling modes. The comparation with the ratios decided by hydrodynamical (hyd.) and rigid (rig.) MOI model are also given.}
\setlength{\tabcolsep}{-0.5mm}
{
\begin{tabular}{c|c|c|c|c|c|c|c|c|c}
\hline
\hline
~~Nucleus~~ & ~\makecell{ Modes \\ in Ref.}~ & ~~~$\gamma$~~~ & ~~~$\beta$~~~ &~~~\makecell{MOI [$\hbar^{2}$MeV$^{-1}$]\\${\cal J}_m,{\cal J}_s,{\cal J}_l$}~~~& ~~\makecell{ ${\cal J}_{m}/{\cal J}_{s}/{\cal J}_{l}$\\ \textrm{PRM}}~~  & ~~\makecell{ ${\cal J}_{m}/{\cal J}_{s}/{\cal J}_{l}$\\ \textrm{ hyd.}}~~ &  ~~\makecell{ ${\cal J}_{m}/{\cal J}_{s}/{\cal J}_{l}$\\ \textrm{rig.}}~~ & ~\makecell{ Behaviour \\of ${\cal J}_{m}/{\cal J}_{s}$} ~ &  ~~Ref.~~\\
\hline

$^{105}$Pd & \red{TW} & $25^{\circ}$ & 0.27 &\makecell{ ${\cal J}_{k}({1+0.023 I(I+1)})^{\frac{1}{2}}$\\ $5.89,3.74,1.27$}  & $1/0.63/0.22$  & $1/0.33/0.18$  & $1/0.89/0.74$  & {Inbetween} & \cite{Pd105}\\

\hline

$^{130}$Ba & \red{TW} & $21.5^{\circ}$ & 0.24 &\makecell{${\cal J}_{k}(1+0.59I)$\\ $1.50,1.09,0.65$}    & $1/0.73/0.43$   & $1/0.40/0.14$  & $1/0.91/0.77$  & {Inbetween} & \cite{Ba130}\\

\hline

$^{133}$La &  LW  &  $26^{\circ}$ & 0.17 & $15.3, 9.1+0.66R, 2.9$    & --   & $1/0.31/0.19$  & $1/0.93/0.83$  & -- & \cite{La133}\\

\hline

\multirow{4}*{$^{135}$Pr} & \red{TW} & $26^{\circ}$ & 0.17 & $21, 12.8+0.14R, 4$    & --   & $1/0.31/0.19$  & $1/0.93/0.83$  & -- & \cite{La133}\\

~ & \red{TW} & $26^{\circ}$ & 0.17 & $21,13,4$    &  $1/0.62/0.19$  & $1/0.31/0.19$  & $1/0.93/0.83$  & {Inbetween} & \cite{Frauendorf14} \\

~ & \red{TW} & $26^{\circ}$ & 0.17 &  \makecell{  ${\cal J}_{k}(1+0.116I),$\\$7.4,5.6,1.8$}    & $1/0.76/0.24$   & $1/0.31/0.19$  & $1/0.93/0.83$  & {Rig.} & \cite{Pr135}\\

~ & Tip & $26^{\circ}$ & 0.17 & \makecell{${\cal J}\sin^{2}(\gamma-2k\pi/3)$\\${\cal J}=12.5\omega(1+\omega^{2})$}    & $1/0.31/0.19$   & $1/0.31/0.19$  & $1/0.93/0.83$  & Hyd. & \cite{Pr135tilted}\\

\hline

$^{161}$Lu & LW & $20^{\circ}$ & 0.42 & $87.56,22.74, 2.77$    & $1/0.26/0.03$   & $1/0.43/0.12$  & $1/0.87/0.62$  & Hyd. & \cite{Raduta2020PRC}\\

\hline

\multirow{2}*{$^{163}$Lu} & \red{TW} & $20^{\circ}$ & 0.42 & $64,56,13$    & $1/0.88/0.20$  & $1/0.43/0.12$  & $1/0.87/0.62$  & {Rig.} & \cite{Frauendorf14}\\

 ~ & LW & $17^{\circ}$ & 0.42 & $63.20,20,10$  & $1/0.32/0.16$   & $1/0.49/0.09$  &     $1/0.89/0.63$  & Hyd. & \cite{Raduta2020PRC}\\

\hline

\multirow{1}*{$^{165}$Lu} & LW & $20^{\circ}$ & 0.42  & $77.30,16.18,4.40$    & $1/0.21/0.06$    & $1/0.43/0.12$  & $1/0.87/0.62$  & Hyd. & \cite{Raduta2020PRC}\\

\hline

\multirow{1}*{$^{167}$Lu} & LW & $19.5^{\circ}$ & 0.43  & $87.03,10.90,3.76$    & $1/0.13/0.04$    & $1/0.44/0.12$  & $1/0.87/0.62$  & Hyd. & \cite{Raduta2020PRC}\\

\hline

\makecell{$^{183}$Au \\ parity+} & \red{TW} & $21.4^{\circ}$ & 0.29 & $50.00,37.52,2.38$   & $1/0.75/0.05$   & $1/0.40/0.14$  & $1/0.90/0.73$  & {Rig.} & \cite{Au183}\\
\hline
\makecell{$^{183}$Au \\ parity-}& \red{TW} & $20^{\circ}$ & 0.30 & $36.85,25.70,5.45$   & $1/0.70/0.15$   & $1/0.43/0.12$  & $1/0.90/0.72$    &{Inbetween} & \cite{Au183}\\

\hline

$^{187}$Au & LW & $23^{\circ}$ & 0.23 & $38\sin^{2}(\gamma-2k\pi/3)$ & $1/0.37/0.15$  & $1/0.37/0.15$  & $1/0.91/0.78$  & Hyd. & \cite{Au187} \\

\hline

\end{tabular}
}
\label{table:MOI}
\end{table}

\clearpage


\begin{thebibliography}{100}
\bibitem{Bohr75}A. Bohr and B. R. Mottelson, \textit{Nuclear Structure}, Vol.II (Benjamin, New York, 1975).
\bibitem{Marshalek79}  E. R. Marshalek, Nucl. Phys. A 331, 429 (1979).
\bibitem{Ru112}J. H. Hamilton, S. J. Zhu, Y. X. Luo, A. V. Ramayya, S. Frauendorf, J.O.Rasmussen \textit{et al}., Nucl. Phys. A {\bf{834}}, 28 (2010).
\bibitem{Lu1631}S. W. {\O}deg{\aa}rd, G. B. Hagemann, D. R. Jensen, M. Bergstr\"{o}m, B. Herskind, G. Sletten \textit{et al}., Phys. Rev. Lett. {\bf{86}}, 5866 (2001).
\bibitem{Lu1632}D. R. Jensen, G. B. Hagemann, I. Hamamoto, S. W. {\O}deg{\aa}rd, B. Herskind, G. Sletten \textit{et al}., Phys. Rev. Lett. {\bf{89}}, 142503 (2002).
\bibitem{Lu161}P. Bringel, G. B. Hagemann, H. H\"{u}bel, A. Al-khatib, P. Bednarczyk, A. B\"{u}rger \textit{et al}., Eur. Phys. J. A {\bf{24}}, 167 (2005).
\bibitem{Lu165}G. Sch\"{o}nwa{\ss}er, H. H\"{u}bel, G. B. Hagemann, P. Bednarczyk,  G. Benzoni, A. Bracco \textit{et al}., Phys. Lett. B {\bf{552}}, 9 (2003).
\bibitem{Lu167}H. Amro, W. C. Ma, G. B. Hagemann, R. M. Diamond, J. Domscheit, P. Fallon \textit{et al}., Phys. Lett. B {\bf{553}}, 197 (2003).
\bibitem{Ta167}D. J. Hartley, R. V. F. Janssens, L. L. Riedinger, M. A. Riley, A. Aguilar, M. P. Carpenter \textit{et al}., Phys. Rev. C {\bf{80}}, 041304 (2009).
\bibitem{Pd105}J. Tim\'{a}r, Q. B. Chen, B. Kruzsicz, D. Sohler, I. Kuti, S. Q. Zhang \textit{et al}., Phys. Rev. Lett. {\bf{122}}, 062501 (2019).
\bibitem{Ba130}Q. B. Chen, S. Frauendorf and C. M. Petrache, Phys. Rev. C {\bf{100}}, 061301(R) (2019).
\bibitem{Ba1302}Y. K. Wang, F. Q. Chen and P. W. Zhao, Phys. Lett. B {\bf{802}}, 135246 (2020).
\bibitem{Ba133} K. Rojeeta Devi, Suresh Kumar, Naveen Kumar, Neelam, F. S. Babra, Md. S. R. Laskar \textit{et al}., Phys. Lett. B {\bf{823}}, 136756 (2021).
\bibitem{Pr135}J. T. Matta, U. Garg, W. Li, S. Frauendorf, A. D. Ayangeakaa, D. Patel \textit{et al}., Phys. Rev. Lett. {\bf{114}}, 082501 (2015).
\bibitem{Pr1352}N. Sensharma, U. Garg, S. Zhu, A. D. Ayangeakaa, S. Frauendorf, W. Li \textit{et al}., Phys. Lett. B  {\bf{792}}, 170 (2019).
\bibitem{Au183}S. Nandi, G. Mukherjee, Q. B. Chen, S. Frauendorf, R. Banik, Soumik Bhattacharya \textit{et al}., Phys. Rev. Lett. {\bf{125}}, 132501 (2020).
\bibitem{Frauendorf14}S. Frauendorf and F. D\"{o}nau, Phys. Rev. C {\bf{89}}, 014322 (2014).
\bibitem{Xe127}S. Chakraborty, H. P. Sharma, S. S. Tiwary, C. Majumder, A. K. Gupta, P. Banerjee \textit{et al}., Phys. Lett. B {\bf{811}}, 135854 (2020).
\bibitem{La133}S. Biswas, R. Palit, S. Frauendorf, U. Garg, W. Li, G. H. Bhat \textit{et al}., Eur. Phys. J. A {\bf{55}}, 159 (2019).
\bibitem{Au187}N. Sensharma, U. Garg, Q. B. Chen, S. Frauendorf, D. P. Burdette, J. L. Cozzi \textit{et al}., Phys. Rev. Lett. {\bf{124}}, 052501 (2020).
\bibitem{Pr135tilted}B. F. Lv, C. M. Petrache, E. A. Lawrie, S. Guo, A. Astier, K. K. Zheng \textit{et al}., Phys. Lett. B {\bf{824}}, 136840 (2022).
\bibitem{Hamamoto02}I. Hamamoto, Phys. Rev. C {\bf{65}}, 044305 (2002).
\bibitem{Hamamoto2003PRC}I. Hamamoto and G. B. Hagemann, Phys. Rev. C {\bf{67}}, 014319 (2003).
\bibitem{Hagemann04}G. B. Hagemann, Eur. Phys. J. A {\bf{20}}, 183 (2003).
\bibitem{Streck18}E. Streck, Q. B. Chen, N. Kaiser, and Ulf-G. Mei{\ss}ner, Phys. Rev. C {\bf{98}}, 044314 (2018).
\bibitem{Broocks21}C. Broocks, Q. B. Chen, N. Kaiser, and Ulf-G. Mei{\ss}ner, Eur. Phys. J. A {\bf{57}}, 161 (2021).
\bibitem{CQB19}Q. B. Chen, S. Frauendorf, N. Kaiser, Ulf-G. Mei{\ss}ner, and J. Meng, Phys. Lett. B {\bf{807}}, 135596 (2020).
\bibitem{Tanabe08}K. Tanabe and K. Sugawara-Tanabe, Phys. Rev. C {\bf{77}}, 064318 (2008).
\bibitem{Tanabe10}K. Sugawara-Tanabe and K. Tanabe, Phys. Rev. C {\bf{82}}, 051303(R) (2010).
\bibitem{Tanabe17}K. Tanabe and K. Sugawara-Tanabe, Phys. Rev. C {\bf{95}}, 064315 (2017).
\bibitem{Budaca2018PRC}R. Budaca, Phys. Rev. C {\bf{97}}, 024302 (2018).
\bibitem{Budaca2021}R. Budaca, Phys. Rev. C {\bf{103}}, 044312 (2021).
\bibitem{Raduta2018}A. A. Raduta, R. Poenaru, and Al. H. Raduta,  J. Phys. G: Nucl. Part. Phys. {\bf{45}}, 105104 (2018).
\bibitem{Raduta2020JPG}A. A. Raduta, R. Poenaru, and C. M. Raduta, J. Phys. G: Nucl. Part. Phys. {\bf{47}}, 025101 (2020).
\bibitem{Raduta2020PRC}A. A. Raduta, R. Poenaru, and C. M. Raduta, Phys. Rev. C {\bf{101}}, 014302 (2020).
\bibitem{Raduta2021}A. A. Raduta, C. M. Raduta, and R. Poenaru, J. Phys. G: Nucl. Part. Phys. {\bf{48}}, 015106 (2021).
\bibitem{Lawrie20}E. A. Lawrie, O. Shirinda, and C. M. Petrache, Phys. Rev. C {\bf{101}}, 034306 (2020).
\bibitem{Matsuzaki02}M. Matsuzaki, Y. R. Shimizu, and K. Matsuyanagi, Phys. Rev. C {\bf{65}}, 041303(R) (2002).
\bibitem{Matsuzaki2003PRC}M. Matsuzaki, Y. R. Shimizu, and K. Matsuyanagi, Eur. Phys. J. A {\bf{20}}, 189 (2003).
\bibitem{Matsuzaki2004PRC}M. Matsuzaki, and S. I. Ohtsubo, Phys. Rev. C {\bf{69}}, 064317 (2004).
\bibitem{Matsuzaki2004PRC_v1}M. Matsuzaki, Y. R. Shimizu, and K. Matsuyanagi, Phys. Rev. C {\bf{69}}, 034325 (2004).
\bibitem{Shimizu2005PRC}Y. R. Shimizu, M. Matsuzaki, and K. Matsuyanagi, Phys. Rev. C {\bf{72}}, 014306 (2005).
\bibitem{Shimizu2008PRC}Y. R. Shimizu, T. Shoji, and M. Matsuzaki, Phys. Rev. C {\bf{77}}, 024319 (2008).
\bibitem{Shoji09}T. Shoji and Y. R. Shimizu, Prog. Theor. Phys. {\bf{121}}, 319 (2009).
\bibitem{Frauendorf2015PRC}S. Frauendorf and F. D\"{o}nau, Phys. Rev. C {\bf{92}}, 064306 (2015).
\bibitem{Nakatsukasa16}T. Nakatsukasa, K. Matsuyanagi, M. Matsuzaki, and Y. R. Shimizu, Phys. Scr. {\bf{91}}, 073008 (2016).
\bibitem{Shimada18}M. Shimada, Y. Fujioka, S. Tagami, and Y. R. Shimizu, Phys. Rev. C {\bf{97}}, 024318 (2018).
\bibitem{CHENQB14}Q. B. Chen, S. Q. Zhang, P. W. Zhao, and J. Meng, Phys. Rev. C {\bf{90}}, 044306 (2014).
\bibitem{CHENQB16}Q. B. Chen, S. Q. Zhang, and J. Meng, Phys. Rev. C {\bf{94}}, 054308 (2016).
\bibitem{WXH18}X. H. Wu, Q. B. Chen, P. W. Zhao, S. Q. Zhang, and J. Meng, Phys. Rev. C {\bf{98}}, 064302 (2018).
\bibitem{Frauendorf182}S. Frauendorf, Phys. Rev. C {\bf{97}}, 069801 (2018).
\bibitem{Tanabe18}K. Tanabe and K. Sugawara-Tanabe, Phys. Rev. C {\bf{97}}, 069802 (2018).
\bibitem{Ring}P. Ring and P. Schuck, \textit{The Nuclear Many Body Problem} (Springer Verlag, Berlin, 1980).
\bibitem{Qi09} B. Qi, S. Q. Zhang, J. Meng and S. Frauendorf, Phys. Lett. B {\bf 675}, 175 (2009).
\bibitem{CFQ17}F. Q. Chen, Q. B. Chen, Y. A. Luo, J. Meng, and S. Q. Zhang, Phys. Rev. C {\bf{96}}, 051303(R) (2017).
\bibitem{CQB18}Q. B. Chen and J. Meng, Phys. Rev. C {\bf{98}}, 031303(R) (2018).
\bibitem{Qi21} B. Qi, H. Zhang, S. Y. Wang, and Q. B. Chen, J. Phys. G {\bf{48}}, 055102 (2021).
\bibitem{Frauendorf20} Q. B. Chen and S. Frauendorf, arXiv:2012.03499 [nucl-th].
\bibitem{Allmond17} J. M. Allmond and J. L. Wood, Phys. Lett. B {\bf 767}, 226 (2017).
\bibitem{Landau} L. D. Landau and E. M. Lifshitz, \textit{Course of Theoretical Physics, Mechanics} (Pergamon Press, London, 1960).
\end{thebibliography}
\end{document}